\documentclass{IEEEtran}
\usepackage{amsmath}
\usepackage{amssymb}
\usepackage{graphicx}
\usepackage{float}
\usepackage{hyperref}
\usepackage[style=ieee, url=false, backend=biber,]{biblatex}
\usepackage{xcolor}
\usepackage{rotating}
\usepackage{pdflscape}
\usepackage{parskip}

\usepackage[symbol]{footmisc}
\graphicspath{ {./images/} }
\usepackage{subcaption}
\usepackage{standalone}

\DeclareBibliographyDriver{article}{%
  \printnames{author} 
  \newunit\newblock
  \printfield{title} 
  \newunit\newblock
  \printfield{shortjournal}
  \setunit{\addspace} 
  \printfield{volume} 
  \setunit*{\addcomma\space} 
  \printfield{pages} 
  \setunit*{\addcomma\space} 
  \usebibmacro{date} 
  \finentry
}

\AtEveryBibitem{%
  \clearname{translator}
  \clearname{number}
  \clearlist{publisher}
  \clearlist{number}
  \clearfield{file}
  \clearfield {pubstate}
  \clearfield {url}
  \clearfield {urldate}
  \clearfield{urlyear}
  \clearfield{urlmonth}
  \clearfield{doi}
  \clearfield{archiveprefix}
  \clearfield{keywords}
  \clearfield{eprint}
}
\begin{document}
\title{Compact Latent Representation for Image Compression (CLRIC)}

\author{\IEEEauthorblockN{Ayman A. Ameen\IEEEauthorrefmark{1} \IEEEauthorrefmark{2},
Thomas Richter\IEEEauthorrefmark{1}, and
André Kaup\IEEEauthorrefmark{3},}

\IEEEauthorblockA{\IEEEauthorrefmark{1}Fraunhofer Institute for Integrated Circuits IIS, Erlangen, Germany}

\IEEEauthorblockA{\IEEEauthorrefmark{2} Department of Physics, Faculty of Science, Sohag University, Egypt}

\IEEEauthorblockA{\IEEEauthorrefmark{3} Friedrich-Alexander University at Erlangen-Nürnberg, Erlangen, Germany}}
\maketitle

\begin{abstract}
Current image compression models often require separate models for each quality level, making them resource-intensive in terms of both training and storage. To address these limitations, we propose an innovative approach that utilizes latent variables from pre-existing trained models (such as the Stable Diffusion Variational Autoencoder) for perceptual image compression. Our method eliminates the need for distinct models dedicated to different quality levels. We employ overfitted learnable functions to compress the latent representation from the target model at any desired quality level. These overfitted functions operate in the latent space, ensuring low computational complexity, around $25.5$ MAC/pixel for a forward pass on images with dimensions $(1363 \times 2048)$ pixels. This approach efficiently utilizes resources during both training and decoding. Our method achieves comparable perceptual quality to state-of-the-art learned image compression models while being both model-agnostic and resolution-agnostic. This opens up new possibilities for the development of innovative image compression methods.
\end{abstract}

\section{Introduction}

Effective compression techniques for images and videos plays a pivotal role for supplying the high quality digital content without the need to update the current infrastructure. Currently, the entertainment services occupy a large portion of the Internet traffic. Such services require high-quality perceptual images and videos and often does not require accurate pixel-wise reconstruction. The current widely used techniques, such as HEVC \cite{sullivanOverviewHighEfficiency2012} and VVC \cite{brossDevelopmentsInternationalVideo2021}, relies on linear operators which are handcrafted to minimize the bitrate of the target. One major issue with these techniques is the production of unacceptable artifacts in the compressed image or video at low bit rate. Learned image compression promises to solve the issues of traditional compression techniques. Currently, there are numerous approaches for learned image compression, such as, generalizable autoencoder and overfitted learnable neural function.  

Recent advancements in learned image compression have demonstrated impressive results in achieving low bit rates \cite{jiangMLICLinearComplexity2024, heELICEfficientLearned2022} and high human visual quality \cite{chenGenerativeVisualCompression2024, liHumanMachineCollaborative2024}, utilizing either generalizable autoencoders or overfitted learnable functions. However, these techniques have some drawbacks and limitations. The generalizable autoencoder approach requires training entirely separate models for image compression, each tailored to specific quality levels with corresponding bit rates, which typically involves training 6-8 distinct models from scratch, each representing a different quality level.

Similarly, the overfitted learnable function method \cite{laduneCOOLCHICCoordinatebasedLow2023, dupontCOINCOmpressionImplicit2021} attempts to fit a single image into a learnable function without leveraging prior information. This approach requires retraining from scratch for every new image and is performed in the image space, which is significantly larger than the latent space.

Our objective is to address these limitations through a novel approach. We trained an overfitted learnable function (such as COOLCHIC) on latents generated by existing latent image models (e.g., a stable diffusion autoencoder), without requiring retraining or backpropagation within the large pretrained latent image model. Our approch yields results comparable to state-of-the-art bit rate and image quality benchmarks, as shown in Figure \ref{fig:Overfitted_on_latent}. The advantages of this novel approach are summarized as follows:
\begin{itemize}

\item \textbf{Utilization of Pre-existing Models:} Leverage existing pretrained image generation models like the stable diffusion autoencoder for image compression without retraining or backpropagation, thus avoiding resource-intensive training processes.

\item \textbf{Efficient Latent Variable Storage:} Efficiently store latent variables from the image generation model, achieving target quality and corresponding bit rates without needing to train a hyper-encoder and decoder.

\item \textbf{Continuous Quality Adaptation:} Achieve continuous quality representation with target bit rates adapted to specific applications through adaptation of the loss function parameters, eliminating the need for separate models for each level of image quality.

\item \textbf{Perceptual Quality Optimization:} Ensure perceptual quality is adaptive to the human visual system, enhancing the viewing experience, as depicted in Figure \ref{fig:combined_assessment_scores}

 \item \textbf{Resource-efficient Overfitting:} Overfit the learnable function in the latent space instead of the image space, reducing memory and computational resource requirements,  since latent space is substantially smaller than image space by a factor of eight or more, enabling tailored image compression for each model using stochastic gradient descent techniques.

\end{itemize}

\section{Related Works}

\subsection{Generalizable Autoencoder}\label{sec:Generalizable_Autoencoder}

Since traditional image compression algorithms such as JPEG 2000 usually utilize only linear decorrelation transformations of the input data, a new neural image compression approach was proposed using autoencoder which is trained on large images or videos dataset. \cite{jiangMLICLinearComplexity2024,jiangMLICMultiReferenceEntropy2024,heELICEfficientLearned2022,theisLossyImageCompression2017a}.  
This approach utilizes the ability of the autoencoder to learn the nonlinear mapping of the images to high-dimensional low entropy latent space and then project them back to the image space \cite{balleEndtoendOptimizedImage2017, balleVariationalImageCompression2018}.  

The autoencoder framework is based on four main components \cite{heELICEfficientLearned2022,jiangMLICLinearComplexity2024}. The first one consisted of analysis neural transfer function $g_a$ to map the image to latent space, followed by a quantization function $Q$ , to quantize the latent. To restore the image back, a synthesis transform $g_s$ is used. One crucial component is a Context-based Entropy model to reduce and estimate the bit-rate. 

The network architecture of the analysis and synthesis transform functions may consist only of convolutional layers \cite{liuComprehensiveBenchmarkSingle2020}. Another architecture introduced residual connection with convolutional layer as base model \cite{liuLearnedImageCompression2023}. Also, transformer-based layers \cite{luTransformerbasedImageCompression2021}, and attention layers mixed with convolutional layers \cite{chengLearnedImageCompression2020} have been utilised as transform functions. 

Since the conditional entropy of the quantized latent vector $\hat{y}$ given a context is smaller than or equal to the entropy of $\hat{y}$ without context, several methods have been proposed for contextual entropy model. A hyper-prior analysis model $h_{a}$ was proposed to extract a context from $\hat{y}$ as $z$, which is quantized to $\hat{z}$. a hyper-prior synthesis model $h_{s}$ creates from $\hat{z}$ parameters of univariate Gaussians which model the probability distribution of $\hat{y}$ for the sake of entropy coding \cite{balleVariationalImageCompression2018}. Other methods have extended the distribution model with Gaussian with mean and scale \cite{minnenJointAutoregressiveHierarchical2018}. Other methods introduced asymmetric Gaussian distribution, Gaussian mixture distribution model \cite{chengLearnedImageCompression2020}, and a Gaussian-Laplacian logistic mixture model \cite{fuLearnedImageCompression2023} as a distribution model that allows more flexibility. 

To improve the compression efficiency, various context-based entropy modeling was proposed. An autoregressive model was introduced that conditioned each pixel with the previously decoded pixels for more effective context modeling \cite{minnenJointAutoregressiveHierarchical2018}. Another context model is the checkerboard convolution which divides the Latent representation into anchor part which is used to extract the context for non-anchor part \cite{chengLearnedImageCompression2020}.  A channel-wise context model \cite{liuUnifiedEndtoEndFramework2020} and channel-wise with unevenly grouped context model \cite{heELICEfficientLearned2022} were introduced to exploit redundancy between channels. Recently, a context model which tried to exploit diverse range of correlations within the latent representation using attention-based architecture \cite{jiangMLICMultiReferenceEntropy2024,jiangMLICLinearComplexity2024}. 

\subsection{Overfitted Neural Function}\label{sec:Overfitted_Neural_Function}

The notion of utilizing an overfitted neural function centers around representing image data as the learnable parameters of a neural function, as opposed to discrete pixel values. This neural function can be evaluated to reconstruct the RGB values of image pixels. The concept gained prominence with the advent of neural radiance fields, a technique for neural rendering \cite{mildenhallNeRFRepresentingScenes2020, mullerInstantNeuralGraphics2022, kerbl3DGaussianSplatting2023}. In the domain of imaging, attempts have been made to represent entire datasets, such as the MNIST dataset, using neural functions to achieve resolution-agnostic representations \cite{kimAttentiveNeuralProcesses2019}. One key advantage of modeling images as neural functions is their resolution agnosticism; images can be represented as continuous neural functions and then evaluated at any desired resolution. This approach assumes that the image signals are inherently continuous.

The pioneering work that introduced the idea of overfitted learnable functions for image compression was COIN \cite{dupontCOINCOmpressionImplicit2021}. COIN employs a simple multilayer perceptron (MLP) to map pixel coordinates to their corresponding $RGB$ values by leveraging the efficiency of periodic activation functions \cite{sitzmannImplicitNeuralRepresentations2020}. Although COIN achieved performance comparable to JPEG compression, it was limited by its inability to exploit pixel locality due to the non-local nature of MLPs. This limitation was mitigated by employing a multi-resolution latent representation followed by a non-linear MLP \cite{mullerInstantNeuralGraphics2022}.

Drawing inspiration from multi-resolution latents, COOL-CHIC introduced an overfitted learned image codec with low decoding complexity that significantly enhanced compression performance compared to COIN.

The COOL-CHIC framework \cite{laduneCOOLCHICCoordinatebasedLow2023, leeEntropyConstrainedImplicitNeural2023, leguayLowComplexityOverfittedNeural2023, blardOverfittedImageCoding2024} consists of four principal components: a multi-resolution latent representation followed by an upsampling kernel, either learned or predefined, subsequently followed by synthesis convolution layers incorporating residual connections. The final component is an autoregressive model that generates a probability distribution for each latent pixel based on previously decoded ones.

\section{Method}
\subsection{Overall Architecture}
\begin{figure}[h]
    \centering
    \includegraphics[width=0.5\textwidth]{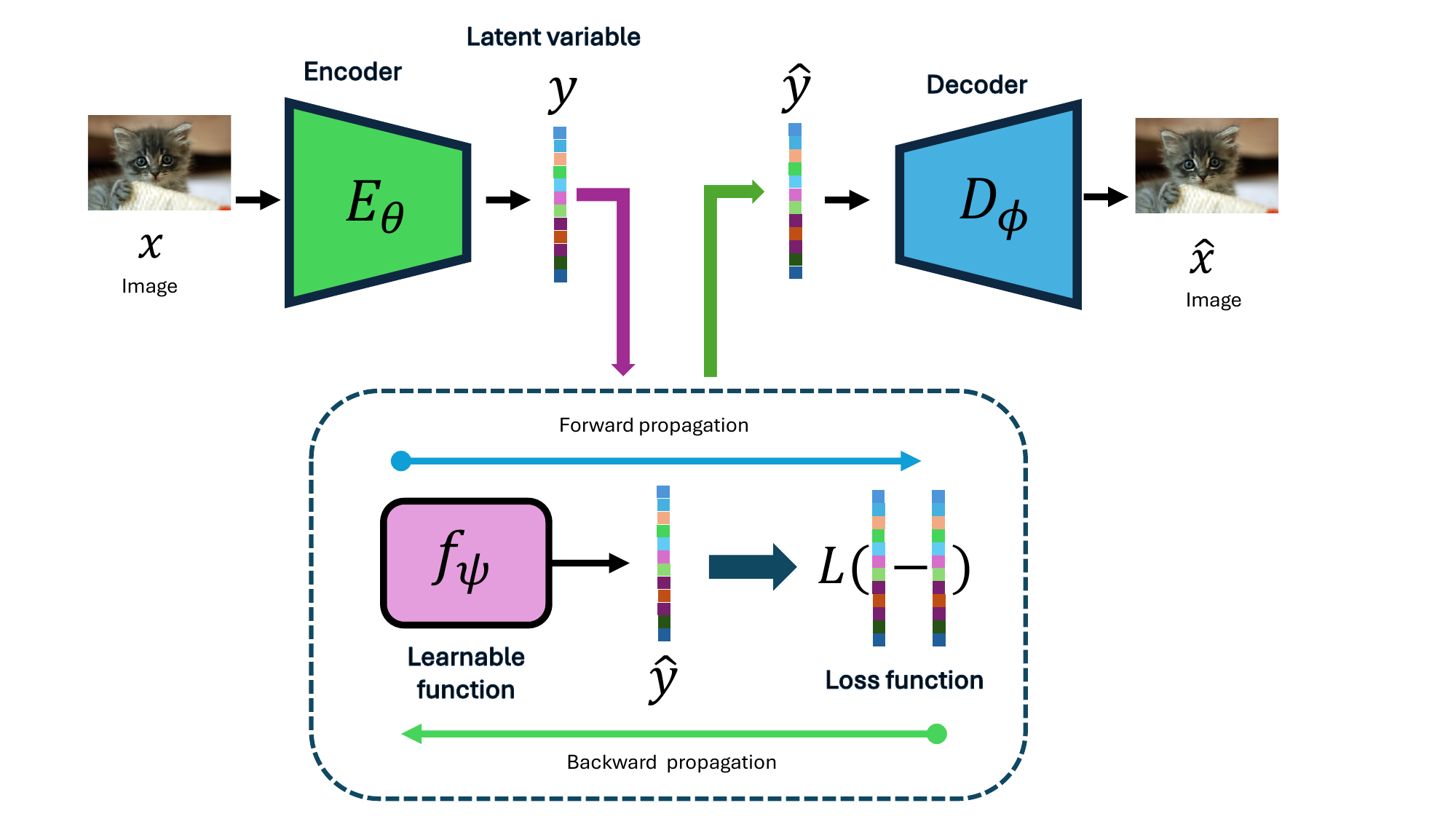}
    \caption{General overview proposed approach of overfitting a learnable function on the latent from generalizable autoencoder}
    \label{fig:Overfitted_on_latent}
\end{figure}

The objective of our work is to introduce an architecture that has the capability representing latent variables from different image models in a compact and efficient manner. 
Our innovative approach leverages compact latent representations from pre-trained models to generate outputs with various bit rates. An overview of our proposed architecture is illustrated in Figure \ref{fig:Overfitted_on_latent}. This architecture allows for mapping any latent variable from a learned image model to a specified bit rate. In this configuration (Figure \ref{fig:Overfitted_on_latent}), an input image $x$ is processed by a pretrained encoder $E_{\theta}$ with parameters $\theta$, resulting in a latent variable $y$. The latent variable serves as desired target output of our learnable function $f_{\psi}$ with parameters $\psi$, which is overfitted to approximate the latent under constraints regarding the target bit rate and its corresponding quality. The loss function is defined as follows \cite{blardOverfittedImageCoding2024,leguayLowComplexityOverfittedNeural2023}:
\begin{equation}\label{equ:loss}
\mathcal{L}(\psi) = D(y, \hat{y}) + \lambda R(\hat{\psi})
\end{equation}

where $R(\hat{\psi})$ represents the rate of the learnable parameters of the overfitted function used to depict the latent $\hat{y}$, $\lambda$ is a Lagrange multiplier that can be tuned to control quality for achieving specific or targeted bit rates, and $D$ is the mean square error in latent Space. Following the training phase, the latent variable $\hat{y}$ is represented by the weights of a learnable function $f_{\psi}$, where $\psi$ denotes the parameters of this function. These parameters are stored or transmitted to the decoder. The function $f_{\psi}$  is evaluated to obtain the latent representation $\hat{y}$ at the decoder side. This latent representation is feeded into a decoder $D$ with pretrained learnable parameters $\phi$, resulting in the reconstructed image $\hat{x}$. 
\subsection{Generative Generalizable Autoencoder}
The generative generalizable autoencoder ($E_\theta, D_\phi$) selected from the Latent Diffusion autoencoder family \cite{rombachHighResolutionImageSynthesis2022, podellSDXLImprovingLatent2023, sauerFastHighResolutionImage2024}, which is pretrained for synthesizing high-resolution images. This selection is based on the models' capability for perceptual compression rather than semantic compression. This capability allows for the generation of images with high perceptual quality at a reduced bit rate. Another advantage of these models is their ability to capture the full data distribution by mapping images from pixel space to latent space with a reduced number of parameters \cite{kimBKSDMLightweightFast2023, chanTutorialDiffusionModels2024, garciaFineTuningImageConditionalDiffusion2024}. This parameter reduction enables the overfitted learnable function to operate effectively with fewer parameters, thereby minimizing training resource requirements. 
\subsection{Overfitted Neural Function}
\begin{figure}[h]
    \centering
    \includegraphics[width=0.5\textwidth]{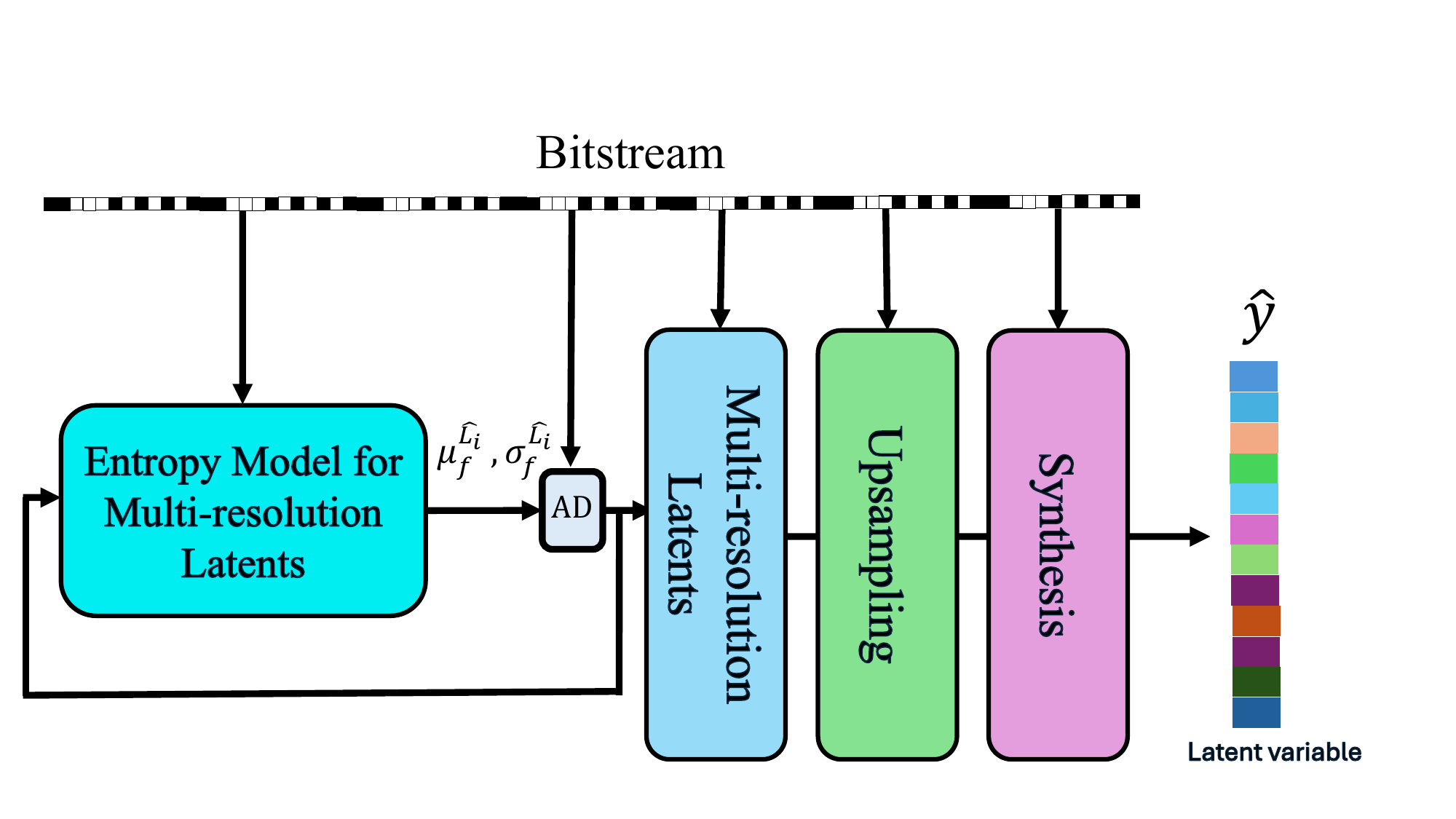}
    \caption{General overview of the proposed overfitted learnable decoder  architecture }
    \label{fig:overfitted_function_latent}
\end{figure}
We utilized COOLCHIC architecture \cite{leguayLowComplexityOverfittedNeural2023,laduneCOOLCHICCoordinatebasedLow2023,blardOverfittedImageCoding2024}  as our overfitted learnable function $f_{\psi}$ as illustrated in Figure \ref{fig:overfitted_function_latent}. 
The proposed architecture consists of three main neural networks \cite{leguayLowComplexityOverfittedNeural2023,laduneCOOLCHICCoordinatebasedLow2023}. The first neural network is an autoregressive model for multi-resolution latents, denoted as $f_{L}$, which provides a probability model characterized by a mean ${\mu}_{f}^{\hat{L_{i}}}$ and a standard deviation ${\sigma}_{f}^{\hat{L_{i}}}$, given the previously decoded latents. The primary objective of the autoregressive process is to enhance the contextual information available for entropy coding. This is because the conditional entropy of $\hat{y}$ given a specific context is less than or equal to the overall entropy of  $\hat{y}$:
$$
H(\hat{y})\leq H(\hat{y}\vert context)
$$
The context can be leveraged to achieve better compression by reducing the bit-rate needed to encode the data. The multi-resolution latents $\hat{\mathbf{L}}$ consist of two-dimensional grids with decreasing size according to:
$$ \hat{\mathbf{L}} = \left\{ \hat{\mathbf{L}}_k \in \mathbb{Z}^{\frac{H}{2^k} \times \frac{W}{2^k}}, k \in 0, \ldots, K-1 \right\}. $$
Here, $H$ and $W$ represent the width and height of a feature map of the latent from the generalization model, and $K$ is the number of the multi-resolution grids.  The second component is an adaptive up-sampling network $f_{u}$, which is initialized with a bicubic sampling kernel and updated during training to adapt to the targeted data \cite{leguayLowComplexityOverfittedNeural2023}. The final network is a synthesis neural network $f_{\psi}$, which produces the final the latent $\hat{y}$ of the generalizable autoencoder. 

\subsection{Effective Bit-Rate Aware Training of Overfitted Neural Function}
The objective of the optimization process is to minimize the distortion between the latent variable $y$ and its reconstructed latent $\hat{y}$, while also reducing the total bit rate associated with all the learnable parameters (latent and model) $\psi$ of the overfit learnable function $f_{\psi}$. This optimization aims to minimize the overall loss, which is expressed in \ref{equ:loss} \cite{laduneCOOLCHICCoordinatebasedLow2023,leguayLowComplexityOverfittedNeural2023,jiangMLICLinearComplexity2024}. The optimization process occurs for every single source image to be encoded as a separate entity.
To reduce the bit rate, quantization plays a crucial role in reducing the bitstream entropy by introducing a lossy compression scheme. However, quantization is a non-differentiable process. To address this issue, we approximate the quantization by adding uniform noise and switch to the actual quantization at the end of the training process of the overfitted neural function $f_{\psi}$ \cite{jiangMLICLinearComplexity2024,jiangMLICMultiReferenceEntropy2024,laduneCOOLCHICCoordinatebasedLow2023}.

\section{Experiments}
\newcommand{\textwidthfigure}{0.8}
\begin{figure*}[h]
\centering
\includegraphics[width=\textwidthfigure\textwidth]{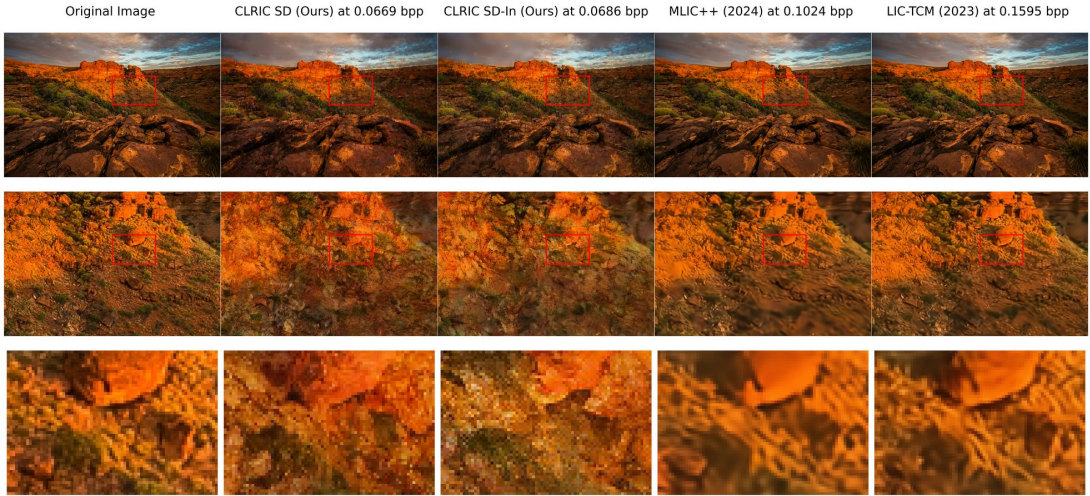}
\captionsetup{width=\textwidthfigure\linewidth}
\caption{Our novel approach performance compared to MLIC++ and LIC-TCM on image num. 27 from the CLIC Professional Valid 2020 dataset.}
\label{fig:0.05_CLIC2020_27_bitrate_zoom_300}
\end{figure*}
\begin{figure*}[h]
\centering
\includegraphics[width=\textwidthfigure\textwidth]{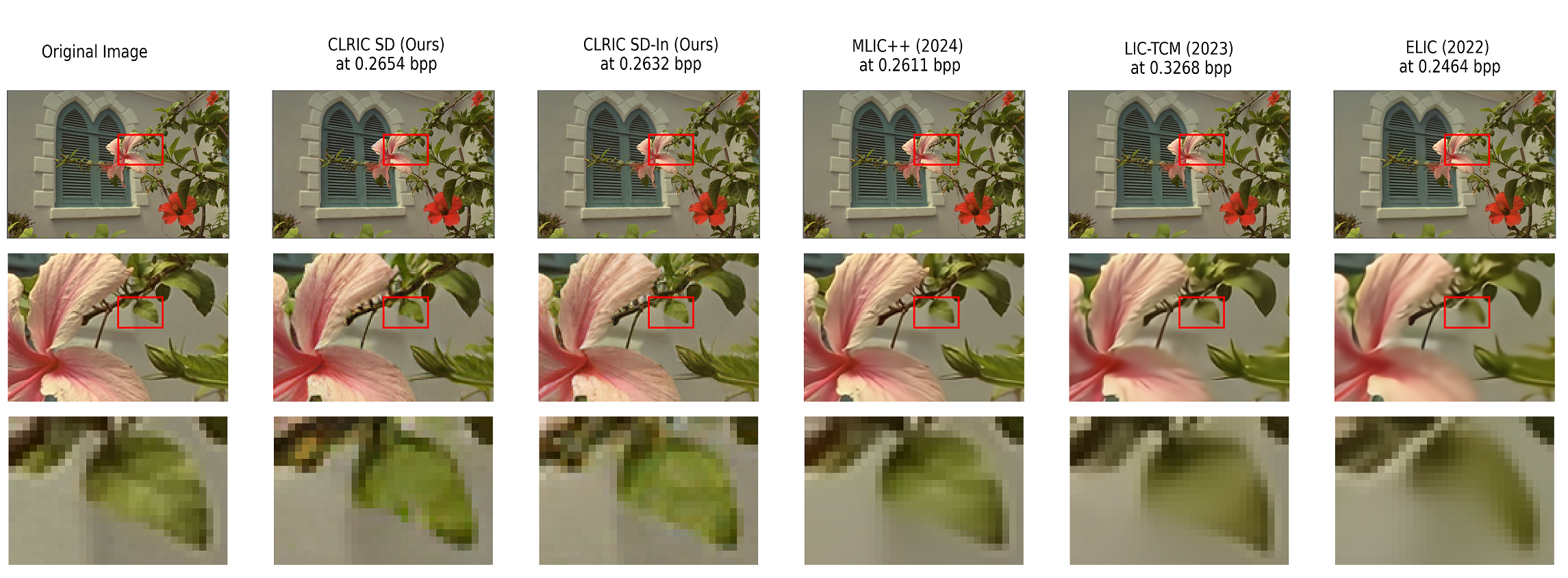}
\captionsetup{width=\textwidthfigure\linewidth}
\caption{Comparison between our approach and different models on image num. 7 from the Kodak dataset.}
\label{fig:0.2_Kodak_7_bitrate_zoom_400}
\end{figure*}
 Initially, we outline our configuration and setup. Subsequently, we evaluate the model's performance through visual results followed by quantitative analysis using perceptual quality metrics and classical distortion metrics. Finally, we demonstrate the capability of continuous quality representation of the latent space at any arbitrary bitrate. Finally, we present our framework's complexity and limitations.
\subsection{Configuration}
This subsection presents the configuration setup for our generalizable autoencoder and our overfitted learnable function. These configurations can be easily modified depending on the target generalizable autoencoder model.
\subsubsection{\textbf{Generalizable Autoencoder}}
We select three derivatives from well-known and widely used open-source pretrained generative autoencoder models. The first model is Stable Diffusion v1.4 autoencoder, referred to in our results as SD \cite{rombachHighResolutionImageSynthesis2022}. The second and third models are Stable Diffusion XL refiner and Stable Diffusion 2 inpainting autoencoders \cite{podellSDXLImprovingLatent2023, sauerFastHighResolutionImage2024}, which we refer to as SD-XL and SD-In, respectively.
\subsubsection{\textbf{Overfitted Learnable Function}}
The overfitted learnable function is based on the COOL-CHIC framework \cite{leguayLowComplexityOverfittedNeural2023,laduneCOOLCHICCoordinatebasedLow2023}, which offers a framework with low decoding complexity. In our experimental setup, we employ an auto-regressive model characterized by an 8-context window. The model outputs the probability distribution for predicting the next latent variable to be decoded. This auto-regressive model comprises three layers with dimensions 8, 8, and 2, respectively, and incorporates a residual connection between the first and second layers. The two-dimensional multi-resolution latent variables are structured into seven grids. The upsampling network utilizes transposed convolutions with a learnable kernel of size 8, initially configured as a bicubic upsampling kernel \cite{leguayLowComplexityOverfittedNeural2023}. The synthesis network is composed of four layers with residual connections every two layers. The output corresponds to the latent size of a Generalizable Autoencoder, with the number of channels matching that of the Generalizable Autoencoder's latent channels. 
\subsubsection{\textbf{Training Strategy}}
The training strategy for this overfitted function is grounded in the COOL-CHIC framework \cite{laduneCOOLCHICCoordinatebasedLow2023,leguayLowComplexityOverfittedNeural2023,blardOverfittedImageCoding2024}, starting with an initial learning rate of $10^{-2}$and employing a decaying learning schedule. Training consists of 15,900 iterations divided into warm-up and main training phases. The warm-up phase includes 2,800 iterations split among five candidates trained over 400 iterations each; the top two candidates undergo an additional 400 iterations. Ultimately, the best candidate proceeds to the main training phase for an additional 13,100 iterations.

\newcommand{\textwidthplot}{0.33}
\begin{figure*}[h]
    \centering
    \begin{subfigure}{\textwidthplot\textwidth}
        \centering
        \includegraphics[width=\textwidth]{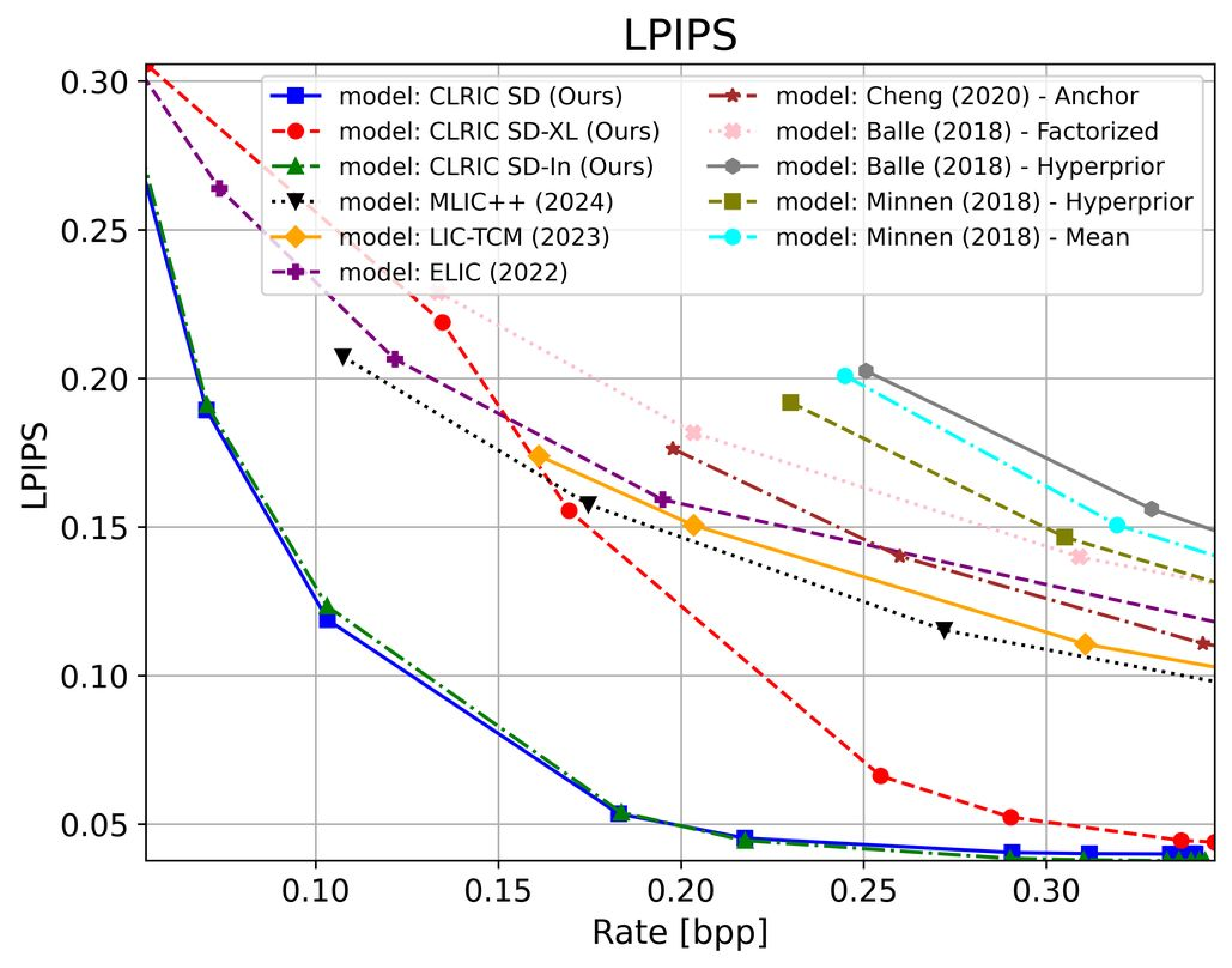}
        \caption{}
        \label{fig:all_LPIPS_KodakDataset}
    \end{subfigure}%
    ~ 
    \begin{subfigure}{\textwidthplot\textwidth}
        \centering
        \includegraphics[width=\textwidth]{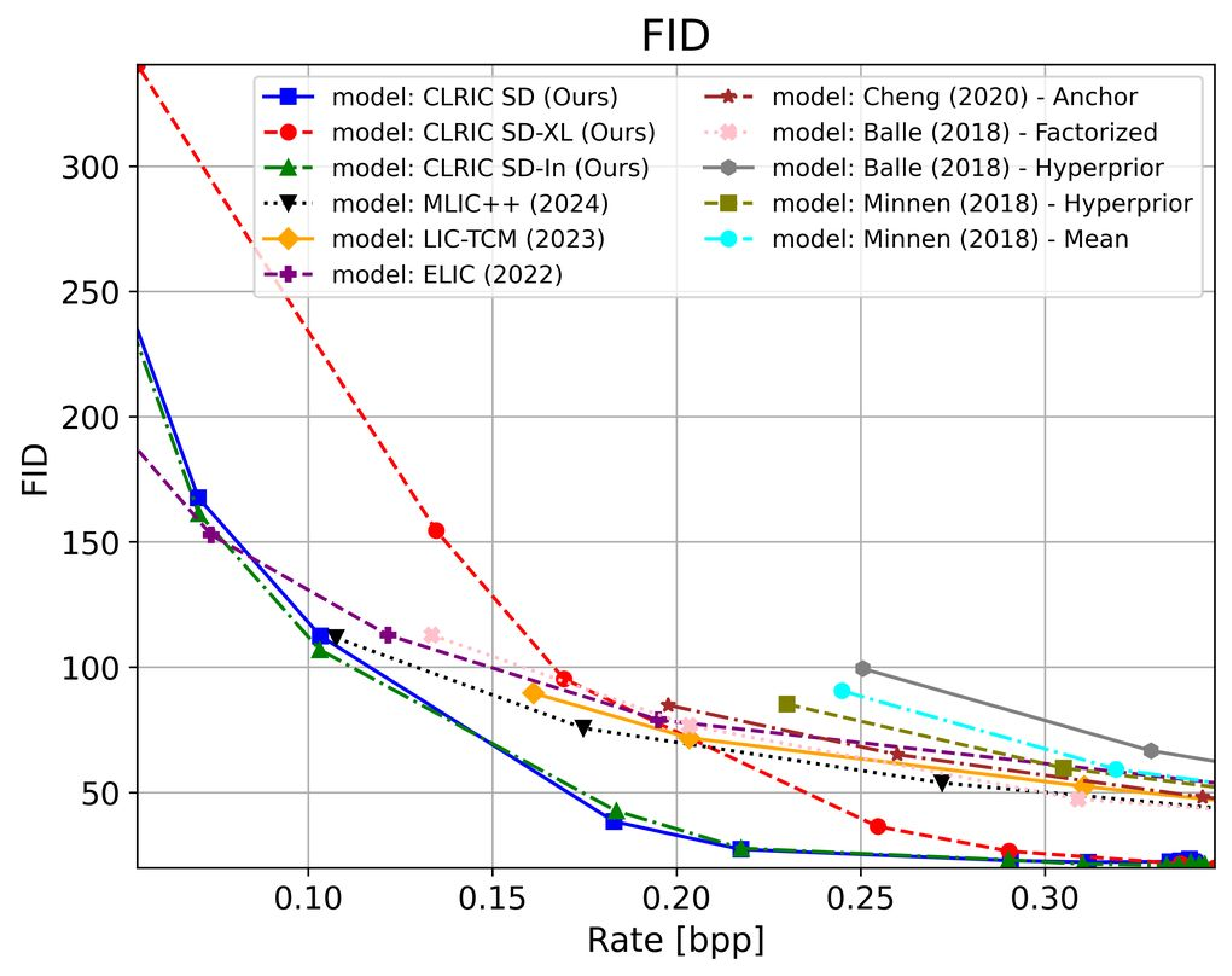}
        \caption{}
        \label{fig:all_FID_KodakDataset}
    \end{subfigure}%
    ~ 
    \begin{subfigure}{\textwidthplot\textwidth}
        \centering
        \includegraphics[width=\textwidth]{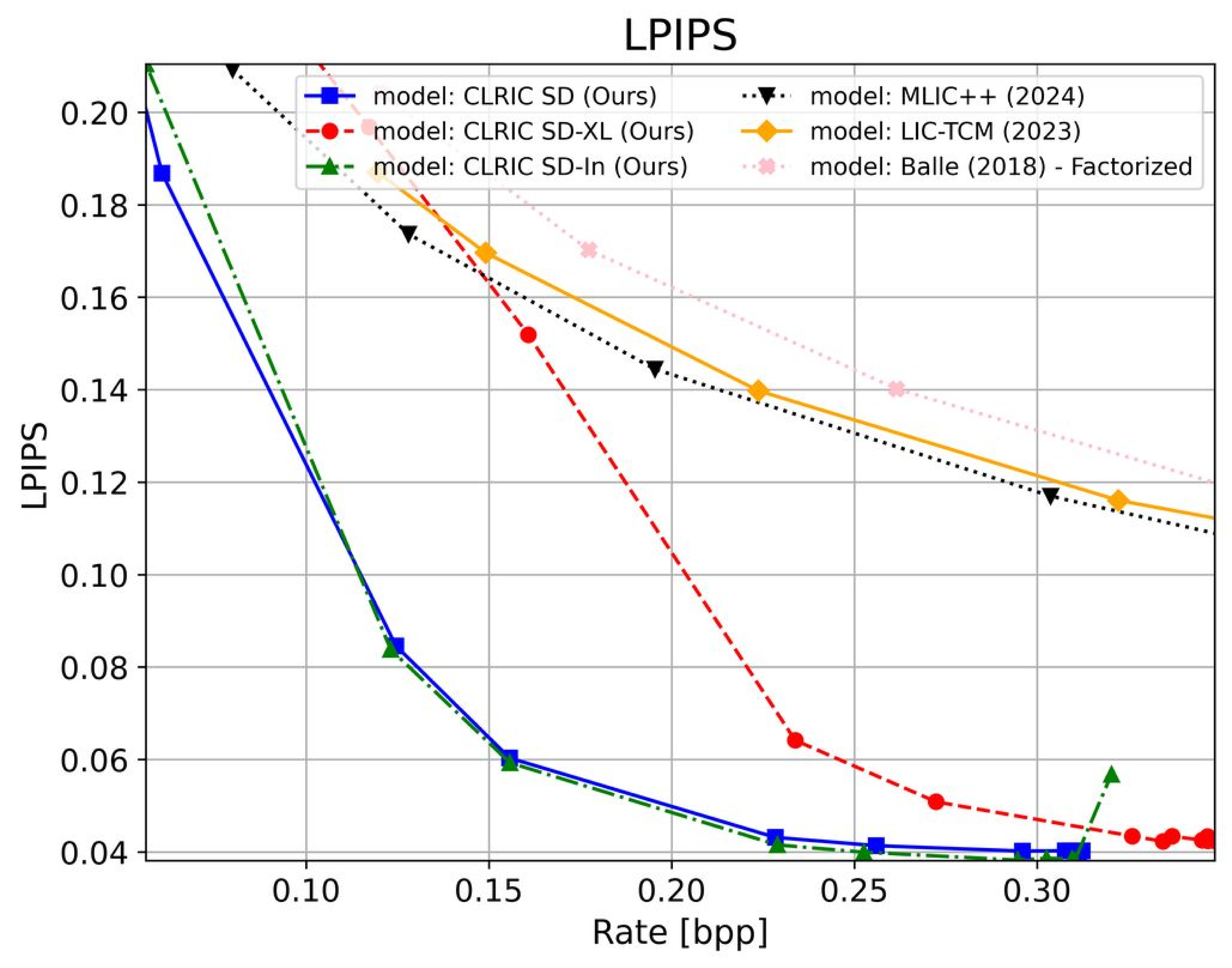}
        \caption{}
        \label{fig:all_LPIPS_professional_valid_2020}
    \end{subfigure}

    \caption{Assessments and comparisons of image compression models using different metrics and datasets. (a) LPIPS scores on the Kodak dataset, (b) FID scores on the Kodak dataset, and (c) LPIPS scores on the CLIC Professional Valid 2020 dataset.}
    \label{fig:combined_assessment_scores} 
\end{figure*}

\subsection{Performance}
To evaluate the performance and generalization capability of our model, we conducted validation using two distinct datasets: the Kodak Dataset and CLIC Professional Valid 2020. The Kodak Dataset, widely used for validating image compression models, consists of 24 images. Meanwhile, CLIC Professional Valid 2020 comprises 41 high-resolution images, making it well-suited for evaluating models in the context of modern digital imagery. We benchmarked our model against these datasets and compared its performance with various other models.
We compared our results to those of several state-of-the-art learned image compression models, including MLIC++ \cite{jiangMLICLinearComplexity2024}, LIC-TCM \cite{liuLearnedImageCompression2023}, and ELIC \cite{heELICEfficientLearned2022}. Additionally, we considered various model variants from notable works such as the Balle (2018) models, specifically the Factorized and Hyperprior variants \cite{balleVariationalImageCompression2018}. We also included the Minnen (2018) model variants, namely Mean and Hyperprior \cite{minnenJointAutoregressiveHierarchical2018}, as well as Cheng (2020) variants, which encompass Anchor and Attention \cite{chengLearnedImageCompression2020}.

\subsubsection{\textbf{Visual Results}}
Our approach employs a Variational Autoencoder (VAE) trained with a perceptual compression strategy, resulting in higher image fidelity and a pleasant perceptual visual experience. This method maintains high granularity and accurately captures the prominent structures in natural images. In contrast, competing models at comparable bitrates often exhibit blurred details or smooth structures with missing granularity.
At low bitrates, where other models tend to produce overly smooth outputs lacking fine details, our method maintains detailed granularity, as shown in Figure \ref{fig:0.05_CLIC2020_27_bitrate_zoom_300}. Our technique effectively reproduces elements like wavy water surfaces and cloudy scenes more accurately compared to other methods that yield overly smooth results.
Our method excels in preserving detail and color fidelity, as illustrated in Figure \ref{fig:0.2_Kodak_7_bitrate_zoom_400}. The synthetic details generated by our model closely resemble the original image's color and texture, whereas other models often suffer from noticeable color shifts and lack of detail. At higher bitrates, our approach retains rich textures that other models' images frequently miss.
\subsubsection{\textbf{Perceptual Quantitative Analysis}}
Our method employs a perceptual image compression strategy, necessitating an analysis of its performance using perceptual metrics. We compared our model against state-of-the-art learned image compression models using the Kodak dataset and the CLIC Professional Validation 2020 dataset (Figure \ref{fig:combined_assessment_scores}). The evaluation focused on perceptual quality, quantified by the Learned Perceptual Image Patch Similarity (LPIPS) metric, in relation to bit rate, denoted in bits per pixel (bpp). LPIPS was selected due to its strong correlation with human visual perception, aligning well with our perceptual image compression approach.
Perceptual quality was assessed for our approach across three different variational autoencoder models. We observed that our approach using the SD and SD-In variational autoencoders as front-end exhibited similar performance levels, with stability diminishing as bit rate increases, as illustrated in Figure \ref{fig:combined_assessment_scores}. Meanwhile, the SD-XL based approach demonstrated slightly worse performance and required higher bit rates to achieve comparable image quality. Other models demanded significantly higher bit rates to match similar levels of perceptual quality.
Furthermore, we utilized the Fréchet Inception Distance (FID) to evaluate image quality. FID measures the similarity between the distributions of reconstructed images and original images. Our method achieved superior results in terms of FID, as depicted in Figure \ref{fig:combined_assessment_scores}. The SD-XL autoencoder-based approach required a higher bit rate to reach an equivalent FID compared to approaches using SD or SD-In.

\section{Conclusion}
In this paper, we presented a novel image compression method based on a two-stages approach. The proposed approach consists of two main components: a generalizable autoencoder capable of leveraging pre-trained generative image models, and an overfitted function that efficiently encodes the latent representation in a compact form. Our method is efficient resource utilization by operating in the latent space rather than the image space. Furthermore, our method allows for continuous quality representation at any target bitrate. We performed analyses using the Learned Perceptual Image Patch Similarity (LPIPS) and Fréchet Inception Distance (FID) metrics. Our approach achieved superior LPIPS and FID scores at comparable bit rates when compared to other models. 

\printbibliography

\onecolumn
\textbf{\large{Supplementary Materials: Compact Latent Representation for Image Compression (CLRIC)}}

\section{Additional Results}

\subsection{\textbf{Classical Distortion Metrics}}

Classical distortion metrics such as Peak Signal-to-Noise Ratio (PSNR), Structural Similarity Index (SSIM), and Multi-Scale SSIM (MS-SSIM) are not ideally suited for evaluating our approach. This is because the variational autoencoders employed in our method are trained using a perceptual image compression strategy \cite{rombachHighResolutionImageSynthesis2022, sauerFastHighResolutionImage2024, podellSDXLImprovingLatent2023}. These autoencoders typically achieve an average PSNR of around 25 dB \cite{rombachHighResolutionImageSynthesis2022}. Meanwhile, our model cannot exceed the PSNR of the autoencoder itself, as it operates on latent representations without accessing the original image data. 

\begin{figure}[h]
    \centering
    \begin{subfigure}{0.45\textwidth}
        \includegraphics[width=\textwidth]{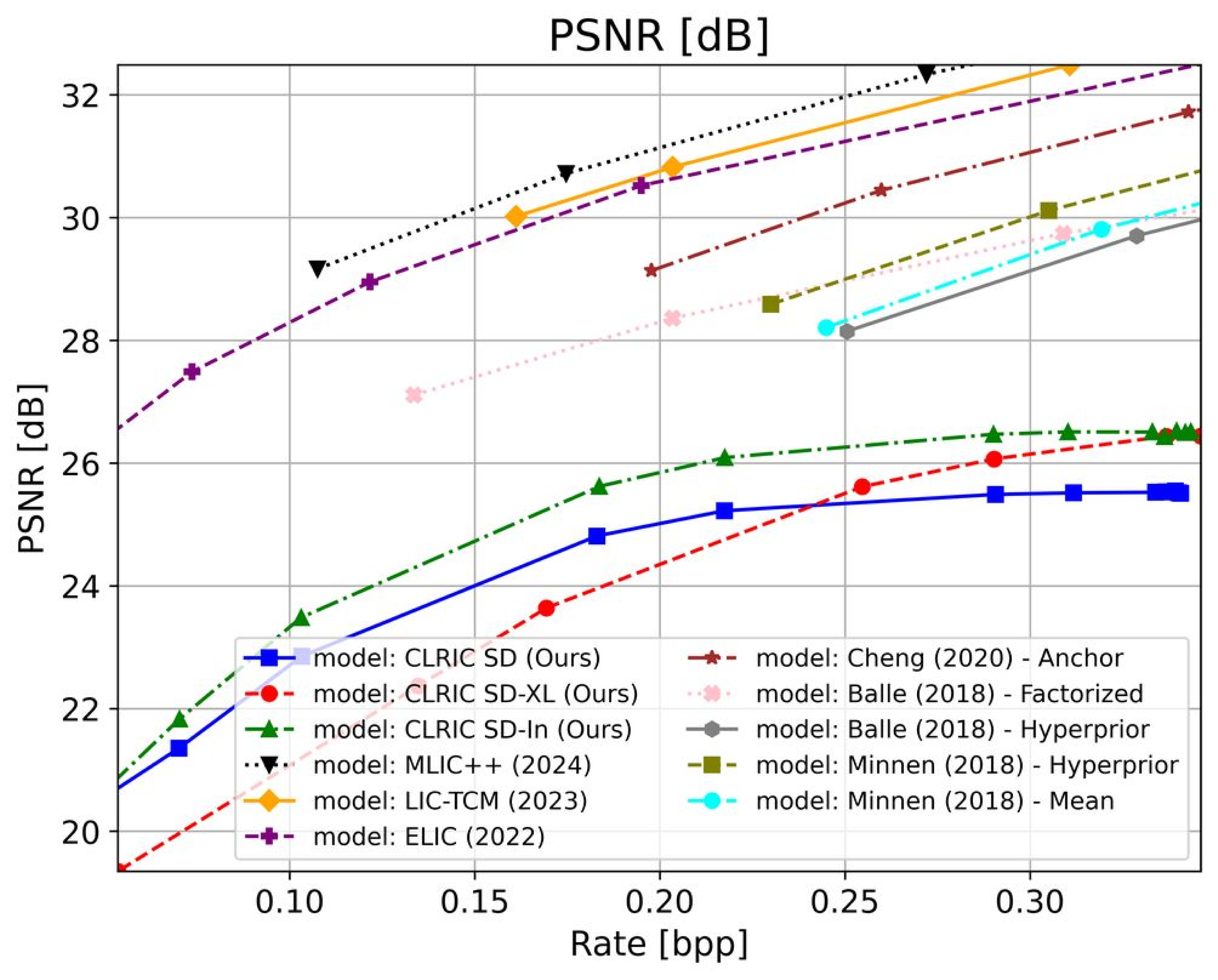}
    \end{subfigure}
    \hfill
    \centering
    \begin{subfigure}{0.45\textwidth}
        \includegraphics[width=\textwidth]{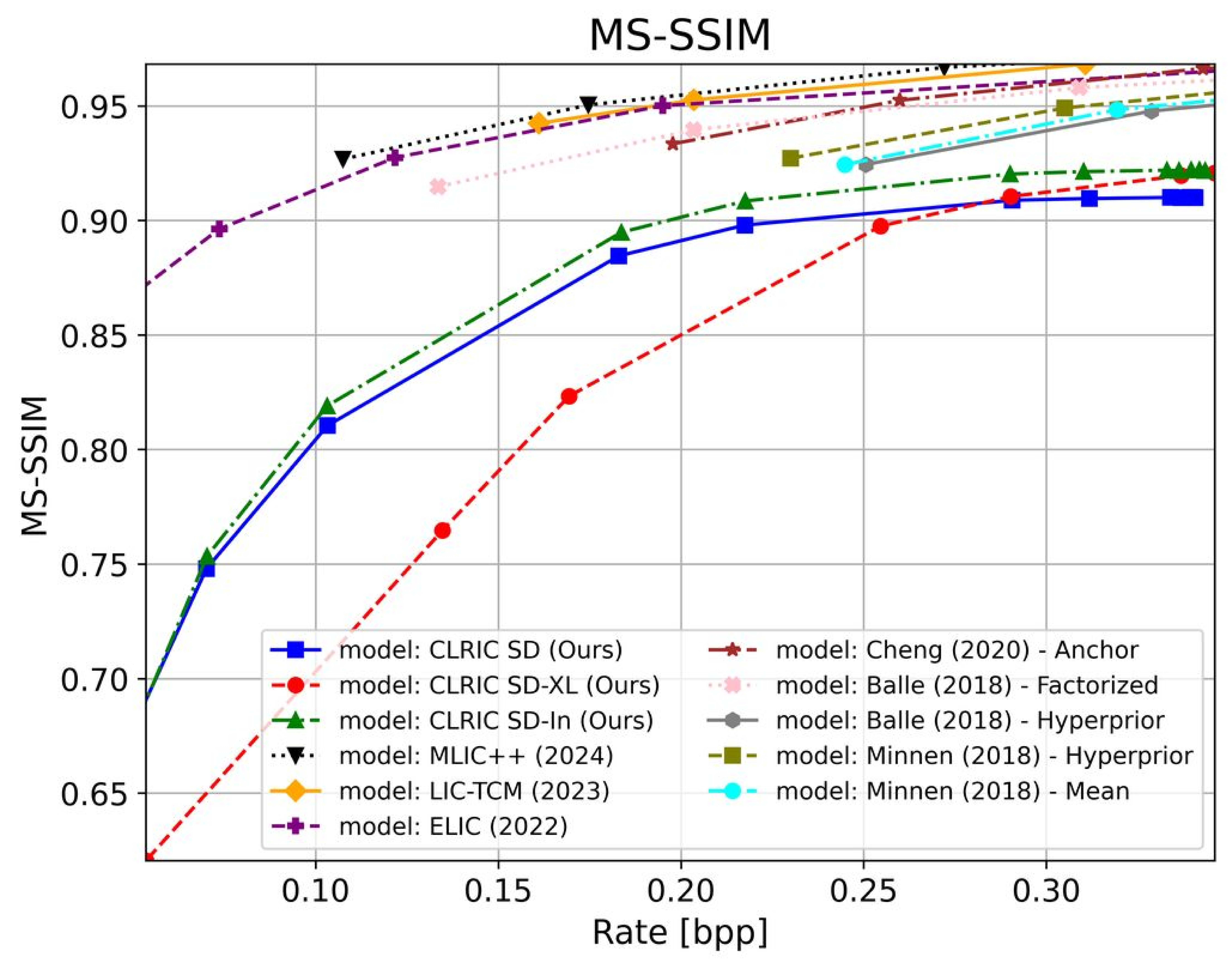}
    \end{subfigure}
    \caption{Comparison of PSNR and MS-SSIM metrics for our method and other learned image compression models on the Kodak dataset.}
    \label{fig:KodakDataset_PSNR_MS_SSIM}
\end{figure}

\begin{figure}[h]
    \centering
    \begin{subfigure}[b]{0.45\textwidth}
        \includegraphics[width=\textwidth]{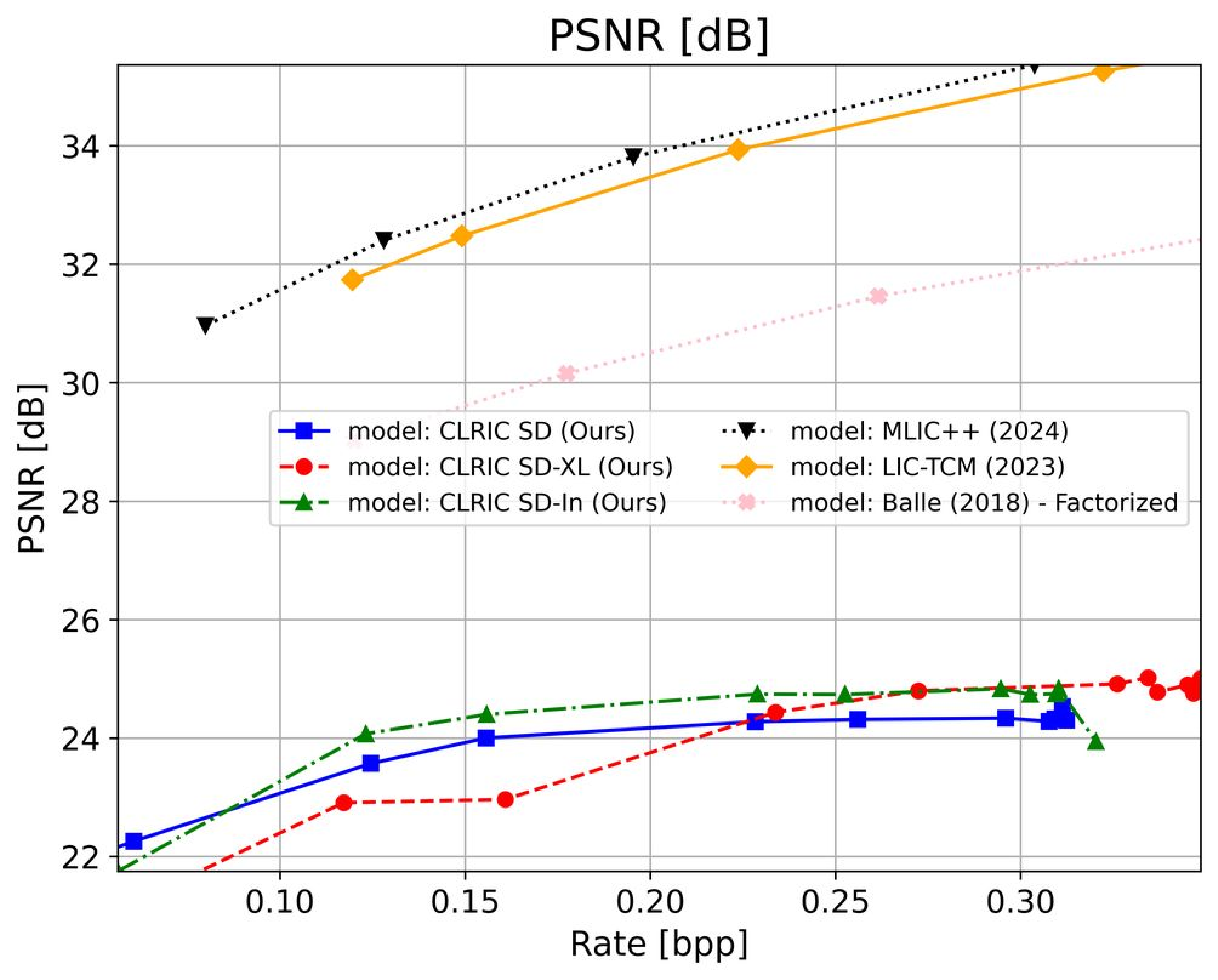}
    \end{subfigure}
    \hfill
    \centering
    \begin{subfigure}[b]{0.45\textwidth}
        \includegraphics[width=\textwidth]{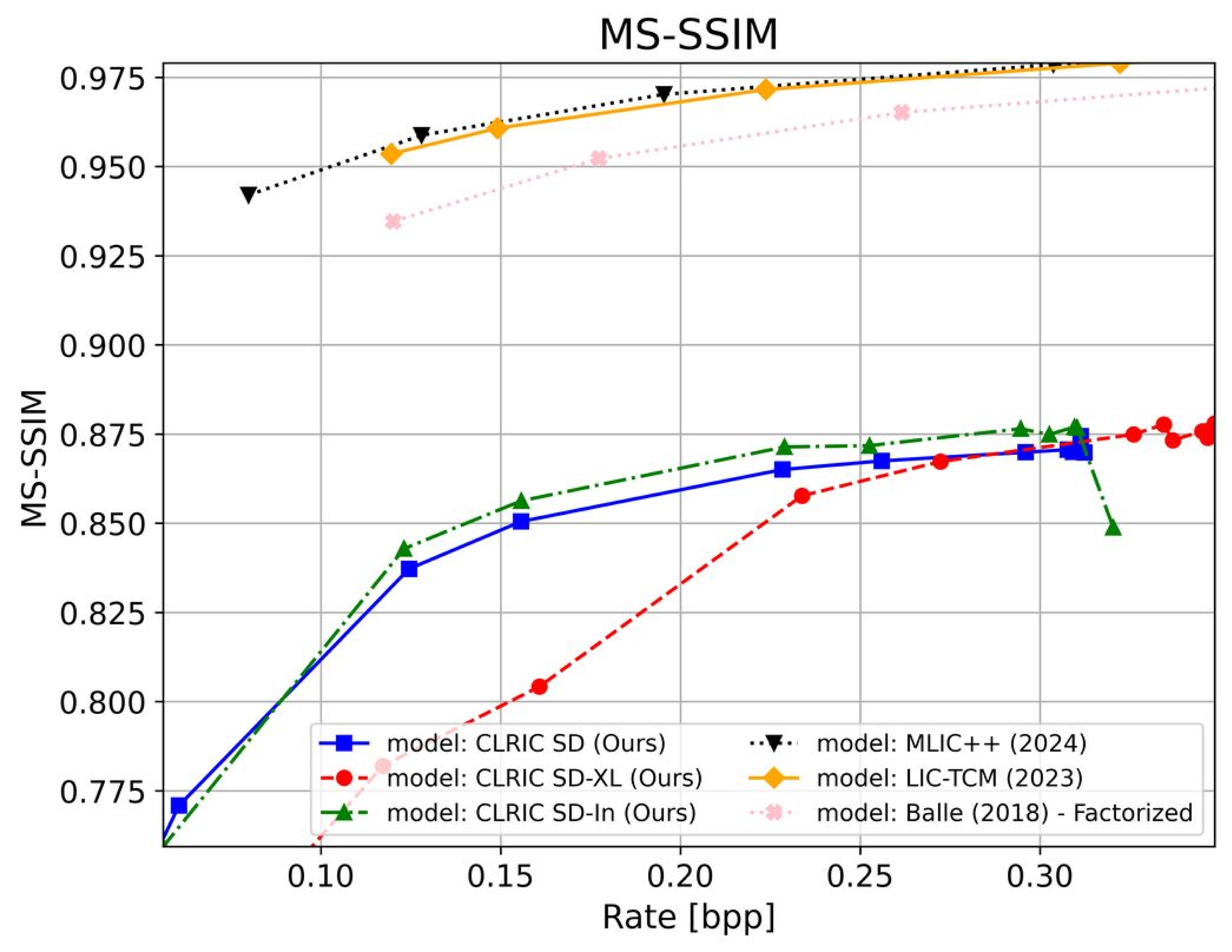}
    \end{subfigure}
    \caption{Evaluation of PSNR and MS-SSIM metrics for our method versus other learned image compression models on the CLIC Professional Valid 2020 dataset.}
    \label{fig:professional_valid_2020_PSNR_MS_SSIM}
\end{figure}

We use classical distortion metrics as a reference point for comparing various learned image codecs. We calculated the average PSNR for both the Kodak Dataset (see Figure \ref{fig:KodakDataset_PSNR_MS_SSIM}) and the CLIC Professional Valid 2020 dataset (see Figure \ref{fig:professional_valid_2020_PSNR_MS_SSIM}). Our approach yields lower PSNR values compared to other learned compression models based on autoencoders, given that PSNR evaluates pixel-wise differences. However, our method provides synthetic textures that closely resemble those in the original images.

The maximum achievable PSNR with our technique is constrained by the maximum PSNR attainable by the original autoencoders (SD, SD-In, and SD-XL) used during training \cite{rombachHighResolutionImageSynthesis2022}.

PSNR increases with higher bitrate; however, it plateaus for models based on SD and SD-In. In contrast, achieving stability requires a higher bitrate when using SD-XL.

A similar trend is observed with MS-SSIM for both the Kodak Dataset and CLIC Professional Valid 2020 dataset (refer to Figures \ref{fig:KodakDataset_PSNR_MS_SSIM} and \ref{fig:professional_valid_2020_PSNR_MS_SSIM}, respectively).

\subsection{\textbf{Continuous Quality Representation}}

\begin{figure}[h] 
    \centering   \includegraphics[width=\textwidth,height=1.9\textheight,keepaspectratio]{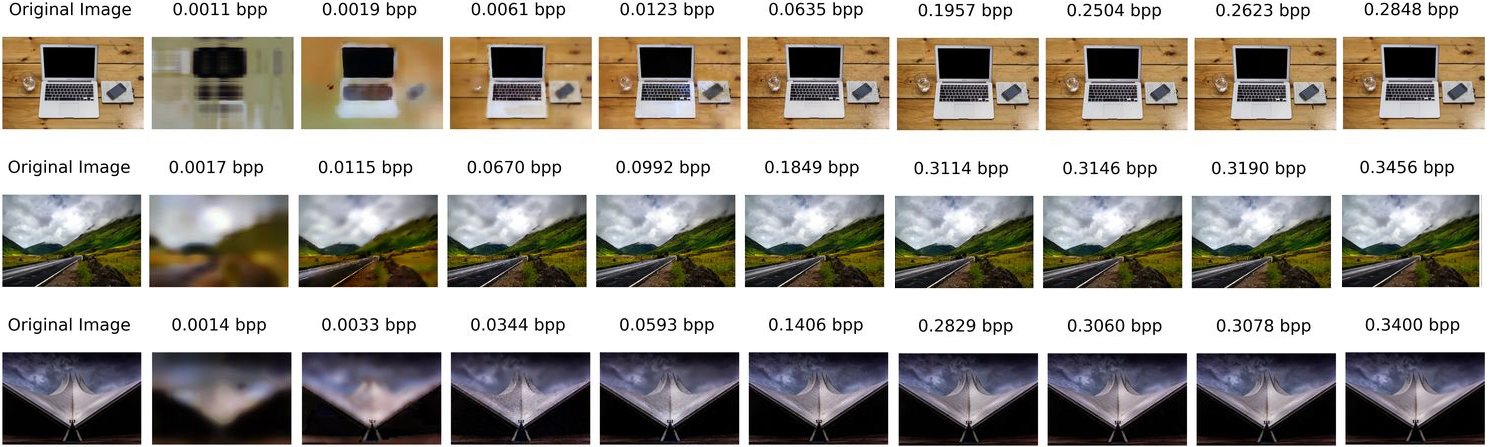}
    \caption{Examples from the CLIC Professional Validation 2020 dataset demonstrating varying bitrates and corresponding quality levels.}
    \label{fig:Continuous_Quality_Representation}
\end{figure}

In contrast to other learned image codecs based on autoencoders with a fixed number of quality levels, our method enables the representation of an image's latent variables at any desired quality level. This ranges from extremely low bitrates, where details are significantly blurred, to higher bitrates that closely resemble the original perceptual quality, as illustrated in \ref{fig:Continuous_Quality_Representation}

At very low bitrates, the image details appear highly blurred, and important structural information is lost. As the bitrate increases, the image quality improves until it becomes perceptually acceptable with enhanced detail.

\subsection{\textbf{Decoding Complexity Analysis of the overfitted neural function}}

In our study, we conducted a decoding complexity analysis of an overfitted neural function trained on latent from SD autoencoder. The results demonstrate that the decoding computational complexity of the overfitted function accounts for approximately 26 multiply-accumulate operations per pixel (MAC/Pixel). This complexity slightly decreases as the image size increases, as detailed in Table \ref{tab:complexity}.

This reduction in the overfitted neural function complexity as the number of pixels increases can be attributed to our approach's operation in latent space rather than directly in image space, which reduces the number of parameters involved.

\begin{table}[H]
    \centering
    \resizebox{\columnwidth}{!}{%
        \begin{tabular}{c c c c c}
            \hline
            Image size & Latent size & Overfitted Fun. (MAC/Pixel) \\ \hline
            512, 768   & 64 , 96     & 25.9              \\ 
            1363, 2048 & 170, 256    & 25.48                       \\ \hline
        \end{tabular}
    }
    \caption{Analysis of decoding complexity of two images with different sizes.}
    \label{tab:complexity}
\end{table}

\subsection{\textbf{Limitations}}

Despite our method achieves state-of-the-art performance in perceptual image quality and low decoding complexity, it presents significant challenges related to encoding complexity and time. Encoding a single image takes approximately 10 minutes on an NVIDIA GeForce GTX 1080 Ti (11 GB) GPU, which limits its practicality for mobile devices. However, this approach could be advantageous for high-demand images or videos (in future work), as the training can be executed once on a server, making evaluation feasible across various devices. Another limitation is the challenge of accurate pixel-wise reconstruction. Although our method excels in perceptual similarity, it is not suitable for compression tasks that require precise pixel-wise fidelity.

\printbibliography

@online{balleEndtoendOptimizedImage2017,
  title = {End-to-End {{Optimized Image Compression}}},
  author = {Ballé, Johannes and Laparra, Valero and Simoncelli, Eero P.},
  date = {2017-03-03},
  eprint = {1611.01704},
  eprinttype = {arXiv},
  eprintclass = {cs, math},
  url = {http://arxiv.org/abs/1611.01704},
  urldate = {2024-03-08},
  pubstate = {prepublished}
}

@online{balleVariationalImageCompression2018,
  title = {Variational Image Compression with a Scale Hyperprior},
  author = {Ballé, Johannes and Minnen, David and Singh, Saurabh and Hwang, Sung Jin and Johnston, Nick},
  date = {2018-05-01},
  eprint = {1802.01436},
  eprinttype = {arXiv},
  eprintclass = {cs, eess, math},
  url = {http://arxiv.org/abs/1802.01436},
  urldate = {2024-08-08},
  langid = {english},
  pubstate = {prepublished}
}

@online{blardOverfittedImageCoding2024,
  title = {Overfitted Image Coding at Reduced Complexity},
  author = {Blard, Théophile and Ladune, Théo and Philippe, Pierrick and Clare, Gordon and Jiang, Xiaoran and Déforges, Olivier},
  date = {2024-03-18},
  eprint = {2403.11651},
  eprinttype = {arXiv},
  doi = {10.48550/arXiv.2403.11651},
  url = {http://arxiv.org/abs/2403.11651},
  urldate = {2024-12-03},
  pubstate = {prepublished}
}

@article{brossDevelopmentsInternationalVideo2021,
  title = {Developments in {{International Video Coding Standardization After AVC}}, {{With}} an {{Overview}} of {{Versatile Video Coding}} ({{VVC}})},
  author = {Bross, Benjamin and Chen, Jianle and Ohm, Jens-Rainer and Sullivan, Gary J. and Wang, Ye-Kui},
  date = {2021-09},
  journaltitle = {Proceedings of the IEEE},
  volume = {109},
  number = {9},
  pages = {1463--1493},
  issn = {1558-2256},
  doi = {10.1109/JPROC.2020.3043399},
  url = {https://ieeexplore.ieee.org/abstract/document/9328514},
  urldate = {2024-11-21},
  eventtitle = {Proceedings of the {{IEEE}}}
}

@online{chanTutorialDiffusionModels2024,
  title = {Tutorial on {{Diffusion Models}} for {{Imaging}} and {{Vision}}},
  author = {Chan, Stanley H.},
  date = {2024-09-06},
  eprint = {2403.18103},
  eprinttype = {arXiv},
  eprintclass = {cs},
  doi = {10.48550/arXiv.2403.18103},
  url = {http://arxiv.org/abs/2403.18103},
  urldate = {2024-09-11},
  pubstate = {prepublished}
}

@online{chenGenerativeVisualCompression2024,
  title = {Generative {{Visual Compression}}: {{A Review}}},
  shorttitle = {Generative {{Visual Compression}}},
  author = {Chen, Bolin and Yin, Shanzhi and Chen, Peilin and Wang, Shiqi and Ye, Yan},
  date = {2024-02-03},
  eprint = {2402.02140},
  eprinttype = {arXiv},
  eprintclass = {cs},
  doi = {10.48550/arXiv.2402.02140},
  url = {http://arxiv.org/abs/2402.02140},
  urldate = {2024-12-23},
  pubstate = {prepublished}
}

@online{chengLearnedImageCompression2020,
  title = {Learned {{Image Compression}} with {{Discretized Gaussian Mixture Likelihoods}} and {{Attention Modules}}},
  author = {Cheng, Zhengxue and Sun, Heming and Takeuchi, Masaru and Katto, Jiro},
  date = {2020-03-30},
  eprint = {2001.01568},
  eprinttype = {arXiv},
  eprintclass = {eess},
  doi = {10.48550/arXiv.2001.01568},
  url = {http://arxiv.org/abs/2001.01568},
  urldate = {2024-08-20},
  pubstate = {prepublished}
}

@online{dupontCOINCOmpressionImplicit2021,
  title = {{{COIN}}: {{COmpression}} with {{Implicit Neural}} Representations},
  shorttitle = {{{COIN}}},
  author = {Dupont, Emilien and Goliński, Adam and Alizadeh, Milad and Teh, Yee Whye and Doucet, Arnaud},
  date = {2021-04-10},
  eprint = {2103.03123},
  eprinttype = {arXiv},
  doi = {10.48550/arXiv.2103.03123},
  url = {http://arxiv.org/abs/2103.03123},
  urldate = {2024-12-02},
  pubstate = {prepublished}
}

@article{fuLearnedImageCompression2023,
  title = {Learned {{Image Compression With Gaussian-Laplacian-Logistic Mixture Model}} and {{Concatenated Residual Modules}}},
  author = {Fu, Haisheng and Liang, Feng and Lin, Jianping and Li, Bing and Akbari, Mohammad and Liang, Jie and Zhang, Guohe and Liu, Dong and Tu, Chengjie and Han, Jingning},
  date = {2023},
  journaltitle = {IEEE Transactions on Image Processing},
  volume = {32},
  pages = {2063--2076},
  issn = {1941-0042},
  doi = {10.1109/TIP.2023.3263099},
  url = {https://ieeexplore.ieee.org/abstract/document/10091784?casa_token=-b_pKCG5l9cAAAAA:kzZhjPCHjBw3OJskhuitArTP0lSKYq5p0DXIeztWmJPhrtVCL4eIk3_PwLXmg2X2ak42ihWt3w},
  urldate = {2024-11-25},
  eventtitle = {{{IEEE Transactions}} on {{Image Processing}}}
}

@online{garciaFineTuningImageConditionalDiffusion2024,
  title = {Fine-{{Tuning Image-Conditional Diffusion Models}} Is {{Easier}} than {{You Think}}},
  author = {Garcia, Gonzalo Martin and Zeid, Karim Abou and Schmidt, Christian and family=Geus, given=Daan, prefix=de, useprefix=false and Hermans, Alexander and Leibe, Bastian},
  date = {2024-09-17},
  eprint = {2409.11355},
  eprinttype = {arXiv},
  doi = {10.48550/arXiv.2409.11355},
  url = {http://arxiv.org/abs/2409.11355},
  urldate = {2024-10-23},
  pubstate = {prepublished}
}

@online{heELICEfficientLearned2022,
  title = {{{ELIC}}: {{Efficient Learned Image Compression}} with {{Unevenly Grouped Space-Channel Contextual Adaptive Coding}}},
  shorttitle = {{{ELIC}}},
  author = {He, Dailan and Yang, Ziming and Peng, Weikun and Ma, Rui and Qin, Hongwei and Wang, Yan},
  date = {2022-03-29},
  eprint = {2203.10886},
  eprinttype = {arXiv},
  eprintclass = {cs, eess},
  doi = {10.48550/arXiv.2203.10886},
  url = {http://arxiv.org/abs/2203.10886},
  urldate = {2024-08-20},
  pubstate = {prepublished}
}

@online{jiangMLICLinearComplexity2024,
  title = {{{MLIC}}++: {{Linear Complexity Multi-Reference Entropy Modeling}} for {{Learned Image Compression}}},
  shorttitle = {{{MLIC}}++},
  author = {Jiang, Wei and Yang, Jiayu and Zhai, Yongqi and Gao, Feng and Wang, Ronggang},
  date = {2024-02-20},
  eprint = {2307.15421},
  eprinttype = {arXiv},
  url = {http://arxiv.org/abs/2307.15421},
  urldate = {2024-11-17},
  pubstate = {prepublished},
  version = {9}
}

@online{jiangMLICMultiReferenceEntropy2024,
  title = {{{MLIC}}: {{Multi-Reference Entropy Model}} for {{Learned Image Compression}}},
  shorttitle = {{{MLIC}}},
  author = {Jiang, Wei and Yang, Jiayu and Zhai, Yongqi and Ning, Peirong and Gao, Feng and Wang, Ronggang},
  date = {2024-01-16},
  eprint = {2211.07273},
  eprinttype = {arXiv},
  url = {http://arxiv.org/abs/2211.07273},
  urldate = {2024-11-17},
  pubstate = {prepublished},
  version = {9}
}

@article{kerbl3DGaussianSplatting2023,
  title = {{{3D Gaussian Splatting}} for {{Real-Time Radiance Field Rendering}}},
  author = {Kerbl, Bernhard and Kopanas, Georgios and Leimkuehler, Thomas and Drettakis, George},
  date = {2023-08},
  journaltitle = {ACM Trans. Graph.},
  volume = {42},
  number = {4},
  pages = {1--14},
  issn = {0730-0301, 1557-7368},
  doi = {10.1145/3592433},
  url = {https://dl.acm.org/doi/10.1145/3592433},
  urldate = {2023-08-01},
  langid = {english}
}

@online{kimAttentiveNeuralProcesses2019,
  title = {Attentive {{Neural Processes}}},
  author = {Kim, Hyunjik and Mnih, Andriy and Schwarz, Jonathan and Garnelo, Marta and Eslami, Ali and Rosenbaum, Dan and Vinyals, Oriol and Teh, Yee Whye},
  date = {2019-07-09},
  eprint = {1901.05761},
  eprinttype = {arXiv},
  doi = {10.48550/arXiv.1901.05761},
  url = {http://arxiv.org/abs/1901.05761},
  urldate = {2024-12-02},
  pubstate = {prepublished}
}

@online{kimBKSDMLightweightFast2023,
  title = {{{BK-SDM}}: {{A Lightweight}}, {{Fast}}, and {{Cheap Version}} of {{Stable Diffusion}}},
  shorttitle = {{{BK-SDM}}},
  author = {Kim, Bo-Kyeong and Song, Hyoung-Kyu and Castells, Thibault and Choi, Shinkook},
  date = {2023-11-16},
  eprint = {2305.15798},
  eprinttype = {arXiv},
  doi = {10.48550/arXiv.2305.15798},
  url = {http://arxiv.org/abs/2305.15798},
  urldate = {2024-10-23},
  pubstate = {prepublished}
}

@inproceedings{laduneCOOLCHICCoordinatebasedLow2023,
  title = {{{COOL-CHIC}}: {{Coordinate-based Low Complexity Hierarchical Image Codec}}},
  shorttitle = {{{COOL-CHIC}}},
  author = {Ladune, Théo and Philippe, Pierrick and Henry, Félix and Clare, Gordon and Leguay, Thomas},
  date = {2023},
  pages = {13515--13522},
  url = {https://openaccess.thecvf.com/content/ICCV2023/html/Ladune_COOL-CHIC_Coordinate-based_Low_Complexity_Hierarchical_Image_Codec_ICCV_2023_paper.html},
  urldate = {2024-08-07},
  eventtitle = {Proceedings of the {{IEEE}}/{{CVF International Conference}} on {{Computer Vision}}},
  langid = {english}
}

@article{leeEntropyConstrainedImplicitNeural2023,
  title = {Entropy-{{Constrained Implicit Neural Representations}} for {{Deep Image Compression}}},
  author = {Lee, Soonbin and Jeong, Jong-Beom and Ryu, Eun-Seok},
  date = {2023},
  journaltitle = {IEEE Signal Processing Letters},
  volume = {30},
  pages = {663--667},
  issn = {1558-2361},
  doi = {10.1109/LSP.2023.3279780},
  url = {https://ieeexplore.ieee.org/document/10132493?denied=},
  urldate = {2024-08-19},
  eventtitle = {{{IEEE Signal Processing Letters}}}
}

@inproceedings{leguayLowComplexityOverfittedNeural2023,
  title = {Low-{{Complexity Overfitted Neural Image Codec}}},
  booktitle = {2023 {{IEEE}} 25th {{International Workshop}} on {{Multimedia Signal Processing}} ({{MMSP}})},
  author = {Leguay, Thomas and Ladune, Théo and Philippe, Pierrick and Clare, Gordon and Henry, Félix and Déforges, Olivier},
  date = {2023-09},
  pages = {1--6},
  issn = {2473-3628},
  doi = {10.1109/MMSP59012.2023.10337636},
  url = {https://ieeexplore.ieee.org/abstract/document/10337636},
  urldate = {2024-08-07},
  eventtitle = {2023 {{IEEE}} 25th {{International Workshop}} on {{Multimedia Signal Processing}} ({{MMSP}})}
}

@article{liHumanMachineCollaborative2024,
  title = {Human–{{Machine Collaborative Image Compression Method Based}} on {{Implicit Neural Representations}}},
  author = {Li, Huanyang and Zhang, Xinfeng},
  date = {2024-06},
  journaltitle = {IEEE Journal on Emerging and Selected Topics in Circuits and Systems},
  volume = {14},
  number = {2},
  pages = {198--208},
  issn = {2156-3365},
  doi = {10.1109/JETCAS.2024.3386639},
  url = {https://ieeexplore.ieee.org/document/10495030?denied=},
  urldate = {2024-08-19},
  eventtitle = {{{IEEE Journal}} on {{Emerging}} and {{Selected Topics}} in {{Circuits}} and {{Systems}}}
}

@article{liuComprehensiveBenchmarkSingle2020,
  title = {A {{Comprehensive Benchmark}} for {{Single Image Compression Artifact Reduction}}},
  author = {Liu, Jiaying and Liu, Dong and Yang, Wenhan and Xia, Sifeng and Zhang, Xiaoshuai and Dai, Yuanying},
  date = {2020},
  journaltitle = {IEEE Transactions on Image Processing},
  volume = {29},
  pages = {7845--7860},
  issn = {1941-0042},
  doi = {10.1109/TIP.2020.3007828},
  url = {https://ieeexplore.ieee.org/abstract/document/9139290?casa_token=rdF1_Gq-J_4AAAAA:LAGd1bVF3cmWFwVo4IG5m3WLBfVxpSGusrnXTQA6TBFpgxS5XgAZfRoaMVczyHIx7f4uZwa3GA},
  urldate = {2024-11-25},
  eventtitle = {{{IEEE Transactions}} on {{Image Processing}}}
}

@inproceedings{liuLearnedImageCompression2023,
  title = {Learned {{Image Compression With Mixed Transformer-CNN Architectures}}},
  author = {Liu, Jinming and Sun, Heming and Katto, Jiro},
  date = {2023},
  pages = {14388--14397},
  url = {https://openaccess.thecvf.com/content/CVPR2023/html/Liu_Learned_Image_Compression_With_Mixed_Transformer-CNN_Architectures_CVPR_2023_paper.html},
  urldate = {2024-11-26},
  eventtitle = {Proceedings of the {{IEEE}}/{{CVF Conference}} on {{Computer Vision}} and {{Pattern Recognition}}},
  langid = {english}
}

@online{liuUnifiedEndtoEndFramework2020,
  title = {A {{Unified End-to-End Framework}} for {{Efficient Deep Image Compression}}},
  author = {Liu, Jiaheng and Lu, Guo and Hu, Zhihao and Xu, Dong},
  date = {2020-05-23},
  eprint = {2002.03370},
  eprinttype = {arXiv},
  doi = {10.48550/arXiv.2002.03370},
  url = {http://arxiv.org/abs/2002.03370},
  urldate = {2024-11-25},
  pubstate = {prepublished}
}

@online{luTransformerbasedImageCompression2021,
  title = {Transformer-Based {{Image Compression}}},
  author = {Lu, Ming and Guo, Peiyao and Shi, Huiqing and Cao, Chuntong and Ma, Zhan},
  date = {2021-11-12},
  eprint = {2111.06707},
  eprinttype = {arXiv},
  doi = {10.48550/arXiv.2111.06707},
  url = {http://arxiv.org/abs/2111.06707},
  urldate = {2024-11-27},
  pubstate = {prepublished}
}

@online{mildenhallNeRFRepresentingScenes2020,
  title = {{{NeRF}}: {{Representing Scenes}} as {{Neural Radiance Fields}} for {{View Synthesis}}},
  shorttitle = {{{NeRF}}},
  author = {Mildenhall, Ben and Srinivasan, Pratul P. and Tancik, Matthew and Barron, Jonathan T. and Ramamoorthi, Ravi and Ng, Ren},
  date = {2020-08-03},
  eprint = {2003.08934},
  eprinttype = {arXiv},
  eprintclass = {cs},
  doi = {10.48550/arXiv.2003.08934},
  url = {http://arxiv.org/abs/2003.08934},
  urldate = {2023-11-07},
  pubstate = {prepublished}
}

@inproceedings{minnenJointAutoregressiveHierarchical2018,
  title = {Joint {{Autoregressive}} and {{Hierarchical Priors}} for {{Learned Image Compression}}},
  booktitle = {Advances in {{Neural Information Processing Systems}}},
  author = {Minnen, David and Ballé, Johannes and Toderici, George D},
  date = {2018},
  volume = {31},
  publisher = {Curran Associates, Inc.},
  url = {https://proceedings.neurips.cc/paper_files/paper/2018/hash/53edebc543333dfbf7c5933af792c9c4-Abstract.html},
  urldate = {2024-08-08}
}

@article{mullerInstantNeuralGraphics2022,
  title = {Instant {{Neural Graphics Primitives}} with a {{Multiresolution Hash Encoding}}},
  author = {Müller, Thomas and Evans, Alex and Schied, Christoph and Keller, Alexander},
  date = {2022-07},
  journaltitle = {ACM Trans. Graph.},
  volume = {41},
  number = {4},
  eprint = {2201.05989},
  eprinttype = {arXiv},
  eprintclass = {cs},
  pages = {1--15},
  issn = {0730-0301, 1557-7368},
  doi = {10.1145/3528223.3530127},
  url = {http://arxiv.org/abs/2201.05989},
  urldate = {2023-09-05},
  langid = {english}
}

@online{podellSDXLImprovingLatent2023,
  title = {{{SDXL}}: {{Improving Latent Diffusion Models}} for {{High-Resolution Image Synthesis}}},
  shorttitle = {{{SDXL}}},
  author = {Podell, Dustin and English, Zion and Lacey, Kyle and Blattmann, Andreas and Dockhorn, Tim and Müller, Jonas and Penna, Joe and Rombach, Robin},
  date = {2023-07-04},
  eprint = {2307.01952},
  eprinttype = {arXiv},
  eprintclass = {cs},
  doi = {10.48550/arXiv.2307.01952},
  url = {http://arxiv.org/abs/2307.01952},
  urldate = {2024-12-24},
  pubstate = {prepublished}
}

@inproceedings{rombachHighResolutionImageSynthesis2022,
  title = {High-{{Resolution Image Synthesis}} with {{Latent Diffusion Models}}},
  booktitle = {2022 {{IEEE}}/{{CVF Conference}} on {{Computer Vision}} and {{Pattern Recognition}} ({{CVPR}})},
  author = {Rombach, Robin and Blattmann, Andreas and Lorenz, Dominik and Esser, Patrick and Ommer, Bjorn},
  date = {2022-06},
  pages = {10674--10685},
  publisher = {IEEE},
  location = {New Orleans, LA, USA},
  doi = {10.1109/CVPR52688.2022.01042},
  url = {https://ieeexplore.ieee.org/document/9878449/},
  urldate = {2023-07-27},
  eventtitle = {2022 {{IEEE}}/{{CVF Conference}} on {{Computer Vision}} and {{Pattern Recognition}} ({{CVPR}})},
  isbn = {978-1-66546-946-3},
  langid = {english}
}

@online{sauerFastHighResolutionImage2024,
  title = {Fast {{High-Resolution Image Synthesis}} with {{Latent Adversarial Diffusion Distillation}}},
  author = {Sauer, Axel and Boesel, Frederic and Dockhorn, Tim and Blattmann, Andreas and Esser, Patrick and Rombach, Robin},
  date = {2024-03-18},
  eprint = {2403.12015},
  eprinttype = {arXiv},
  eprintclass = {cs},
  doi = {10.48550/arXiv.2403.12015},
  url = {http://arxiv.org/abs/2403.12015},
  urldate = {2024-12-24},
  pubstate = {prepublished}
}

@inproceedings{sitzmannImplicitNeuralRepresentations2020,
  title = {Implicit {{Neural Representations}} with {{Periodic Activation Functions}}},
  booktitle = {Advances in {{Neural Information Processing Systems}}},
  author = {Sitzmann, Vincent and Martel, Julien and Bergman, Alexander and Lindell, David and Wetzstein, Gordon},
  date = {2020},
  volume = {33},
  pages = {7462--7473},
  publisher = {Curran Associates, Inc.},
  url = {https://proceedings.neurips.cc/paper/2020/hash/53c04118df112c13a8c34b38343b9c10-Abstract.html},
  urldate = {2024-12-03}
}

@article{sullivanOverviewHighEfficiency2012,
  title = {Overview of the {{High Efficiency Video Coding}} ({{HEVC}}) {{Standard}}},
  author = {Sullivan, Gary J. and Ohm, Jens-Rainer and Han, Woo-Jin and Wiegand, Thomas},
  date = {2012-12},
  journaltitle = {IEEE Transactions on Circuits and Systems for Video Technology},
  volume = {22},
  number = {12},
  pages = {1649--1668},
  issn = {1558-2205},
  doi = {10.1109/TCSVT.2012.2221191},
  url = {https://ieeexplore.ieee.org/abstract/document/6316136},
  urldate = {2024-11-21},
  eventtitle = {{{IEEE Transactions}} on {{Circuits}} and {{Systems}} for {{Video Technology}}}
}

@online{theisLossyImageCompression2017a,
  title = {Lossy {{Image Compression}} with {{Compressive Autoencoders}}},
  author = {Theis, Lucas and Shi, Wenzhe and Cunningham, Andrew and Huszár, Ferenc},
  date = {2017-03-01},
  eprint = {1703.00395},
  eprinttype = {arXiv},
  eprintclass = {cs, stat},
  doi = {10.48550/arXiv.1703.00395},
  url = {http://arxiv.org/abs/1703.00395},
  urldate = {2024-08-08},
  pubstate = {prepublished}
}

\section{Additional Figures}
\begin{figure}[ht]
\centering
\includegraphics[width=\textwidth]{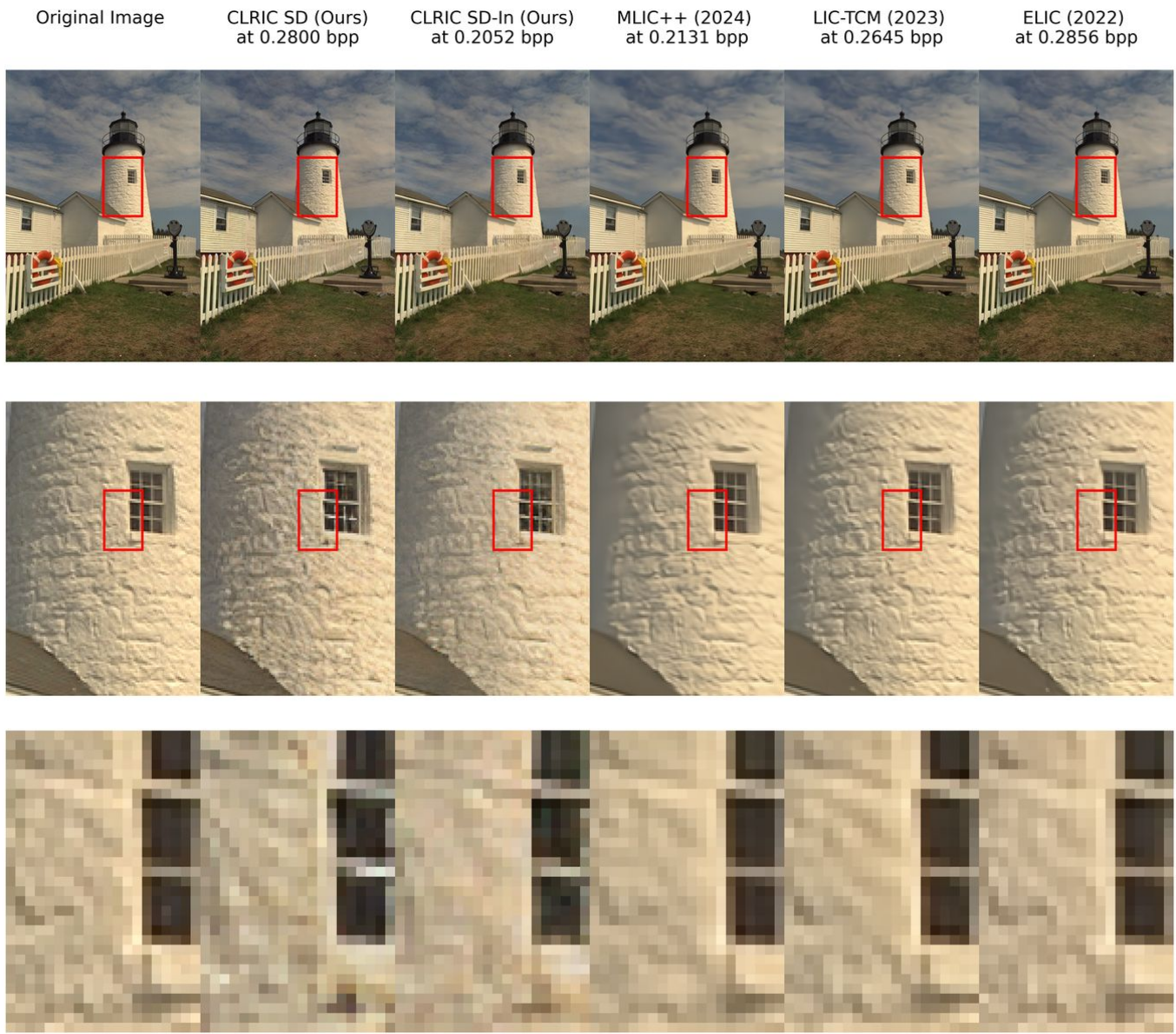}
\caption{Comparison of our approach against various models on image number 19 from the Kodak dataset.}
\label{fig:0.2_Kodak_19_bitrate_zoom_400}
\end{figure}

\begin{figure*}[ht]
\centering
\includegraphics[width=\textwidth]{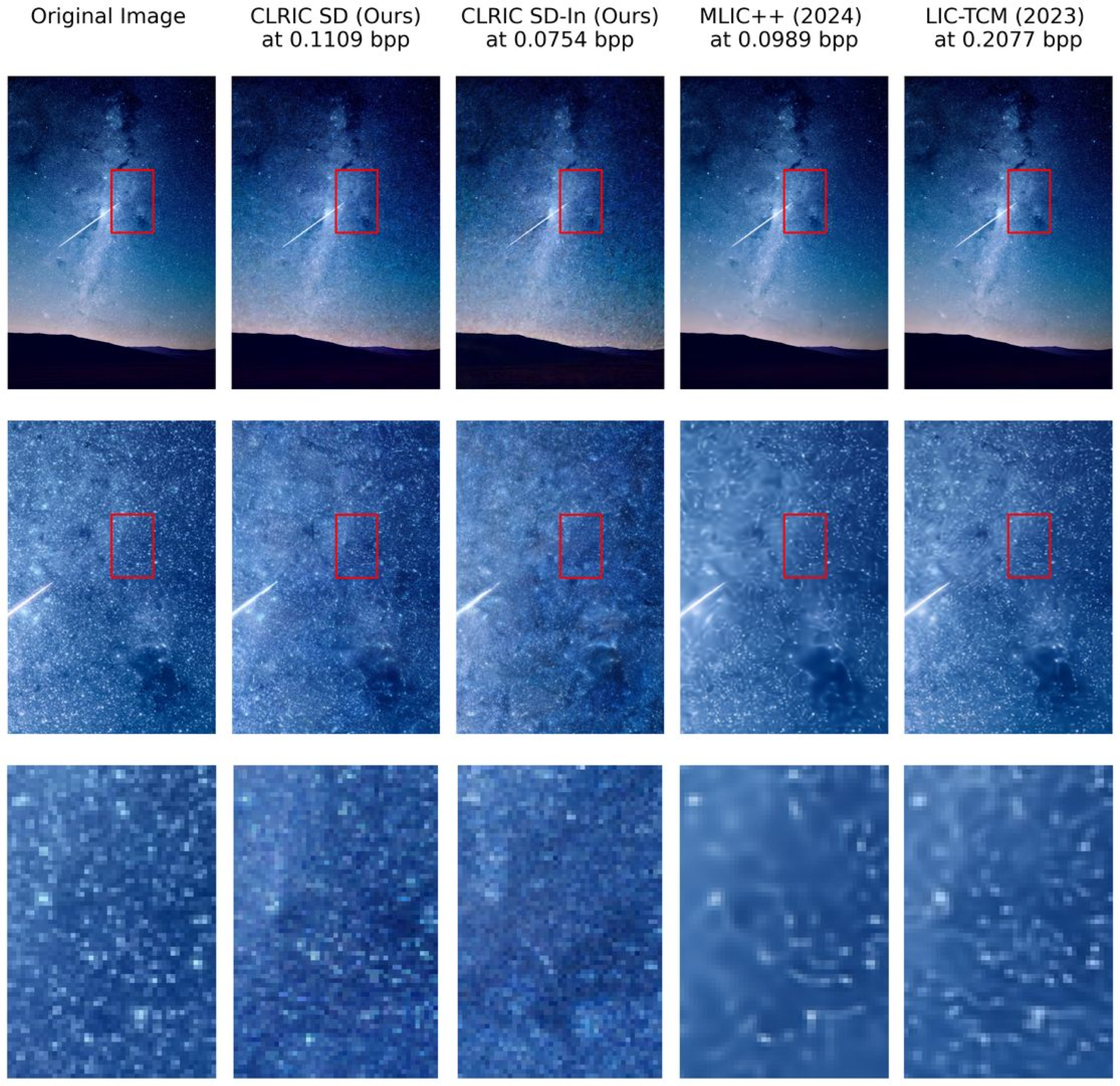}
\caption{Comparison of our approach against various models on image number 18 from the CLIC Professional Valid 2020 dataset.}
\label{fig:0.05_CLIC2020_18_bitrate_zoom_300}
\end{figure*}

\begin{figure*}[ht]
\centering
\includegraphics[width=\textwidth]{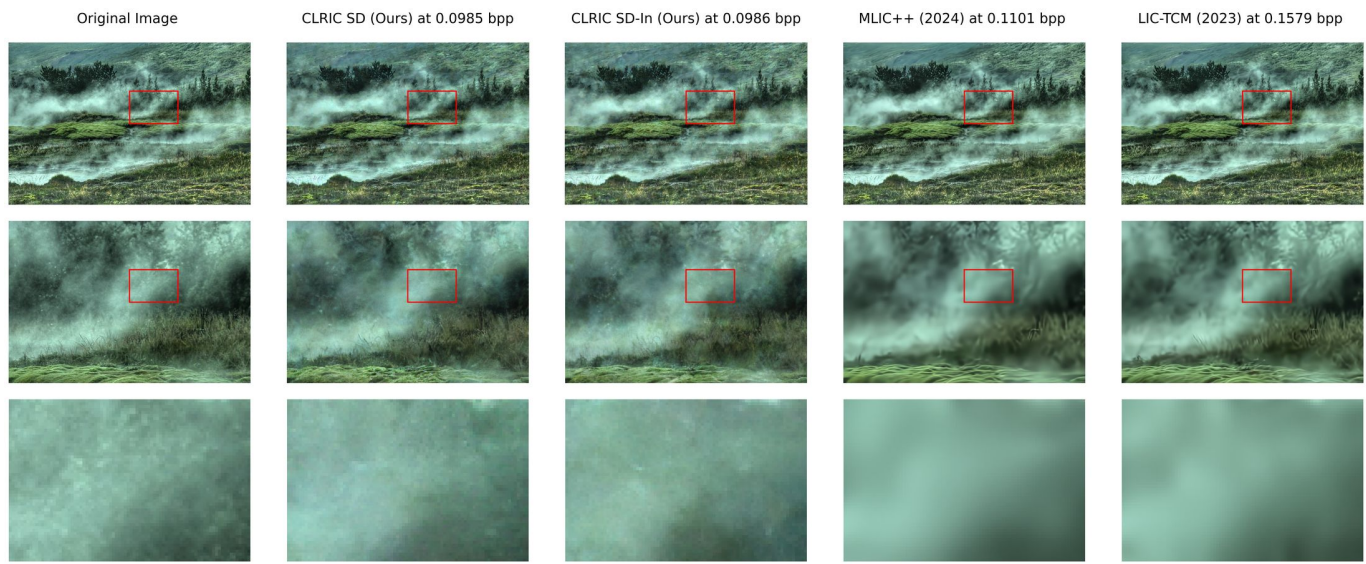}
\caption{Comparison of our approach against various models on image number 14 from the CLIC Professional Valid 2020 dataset.}
\label{fig:0.05_CLIC2020_14_bitrate_zoom_300}
\end{figure*}

\begin{figure*}[ht]
\centering
\includegraphics[width=\textwidth]{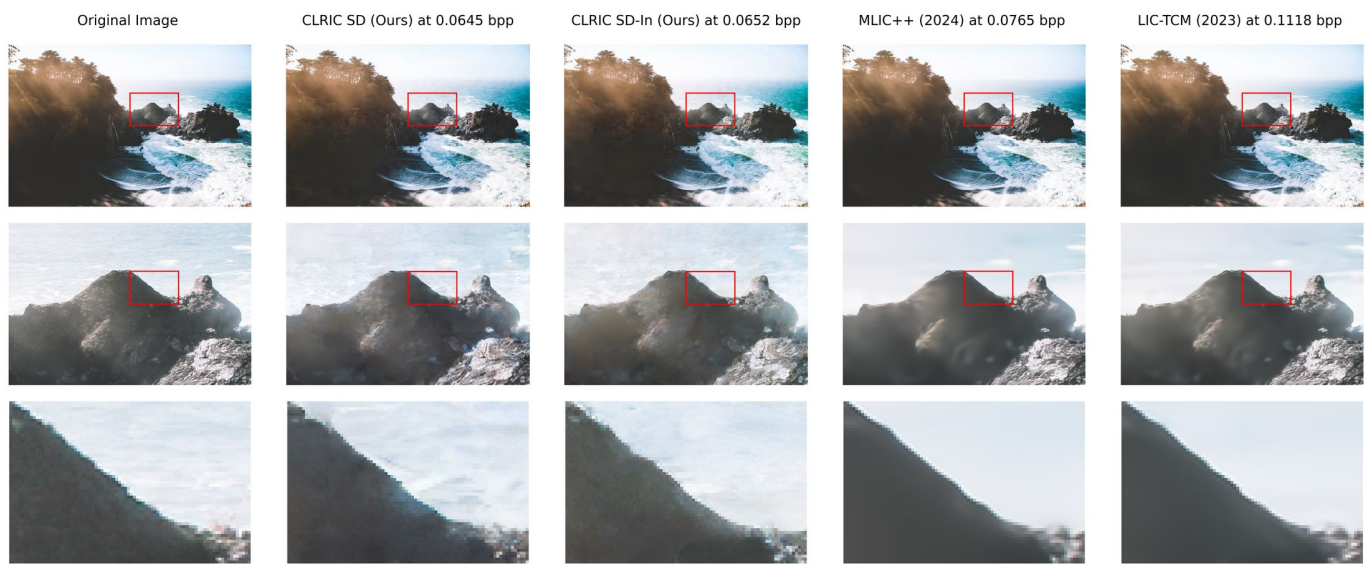}
\caption{Comparison of our approach against various models on image number 23 from the CLIC Professional Valid 2020 dataset.}
\label{fig:0.05_CLIC2020_23_bitrate_zoom_300}
\end{figure*}

\begin{figure*}[ht]
\centering
\includegraphics[width=\textwidth]{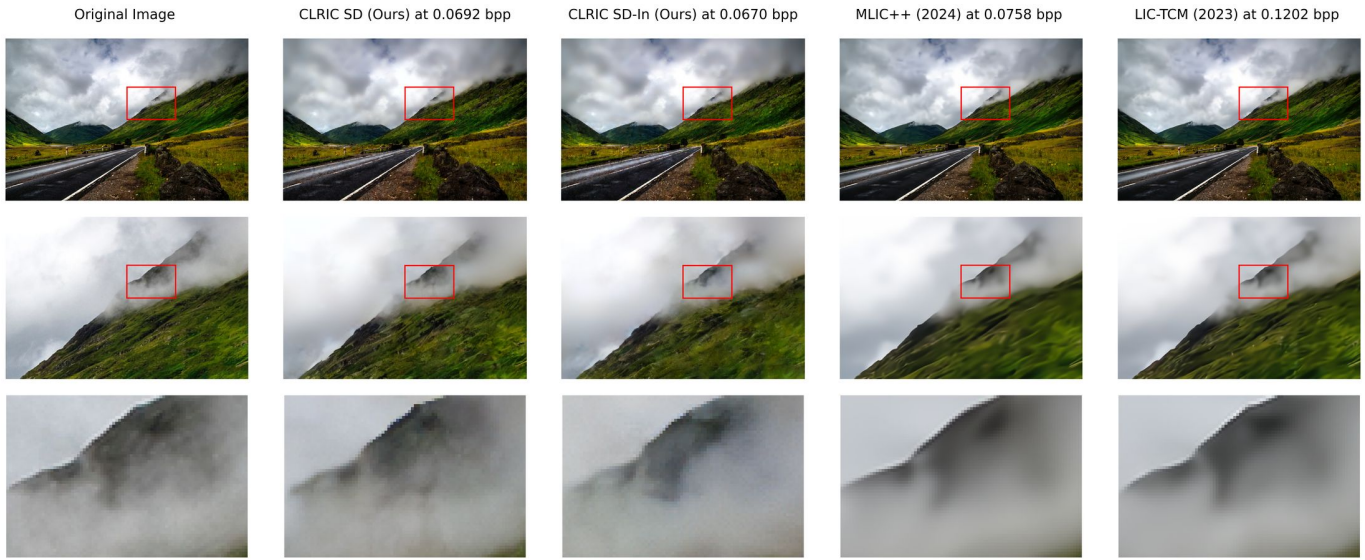}
\caption{Comparison of our approach against various models on image number 7 from the CLIC Professional Valid 2020 dataset.}
\label{fig:0.05_CLIC2020_7_bitrate_zoom_300}
\end{figure*}

\begin{figure*}[ht]
\centering
\includegraphics[width=\textwidth]{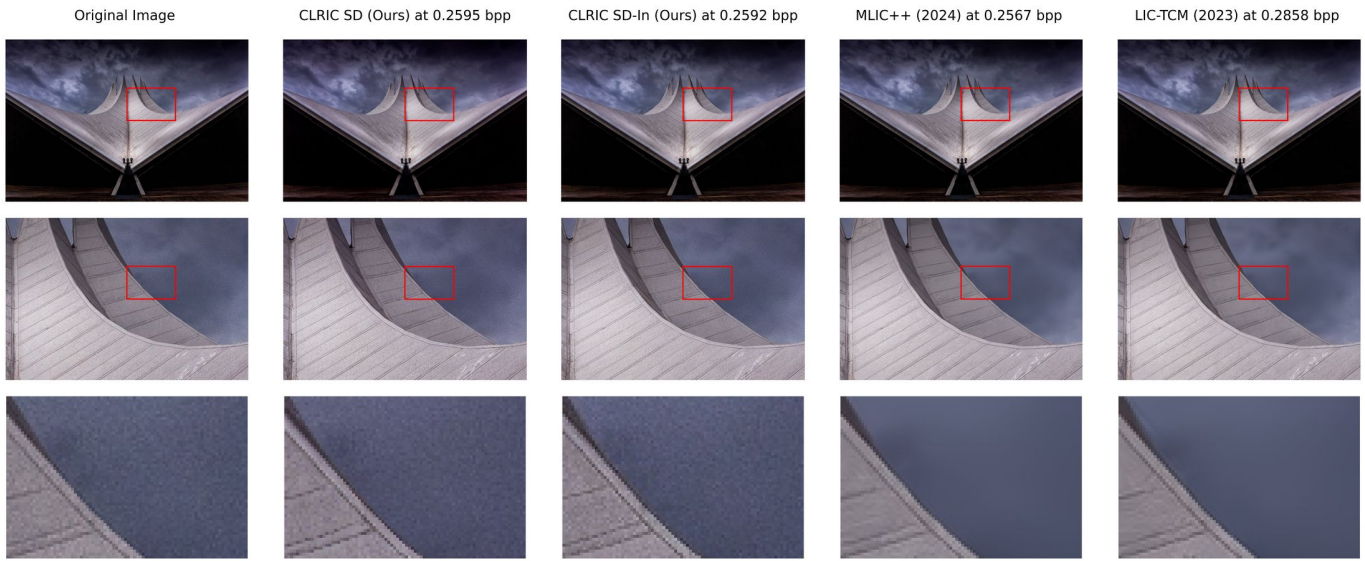}
\caption{Comparison of our approach against various models on image number 20 from the CLIC Professional Valid 2020 dataset.}
\label{fig:0.2_CLIC2020_20_bitrate_zoom_300}
\end{figure*}

\begin{figure*}[ht]
\centering
\includegraphics[width=\textwidth]{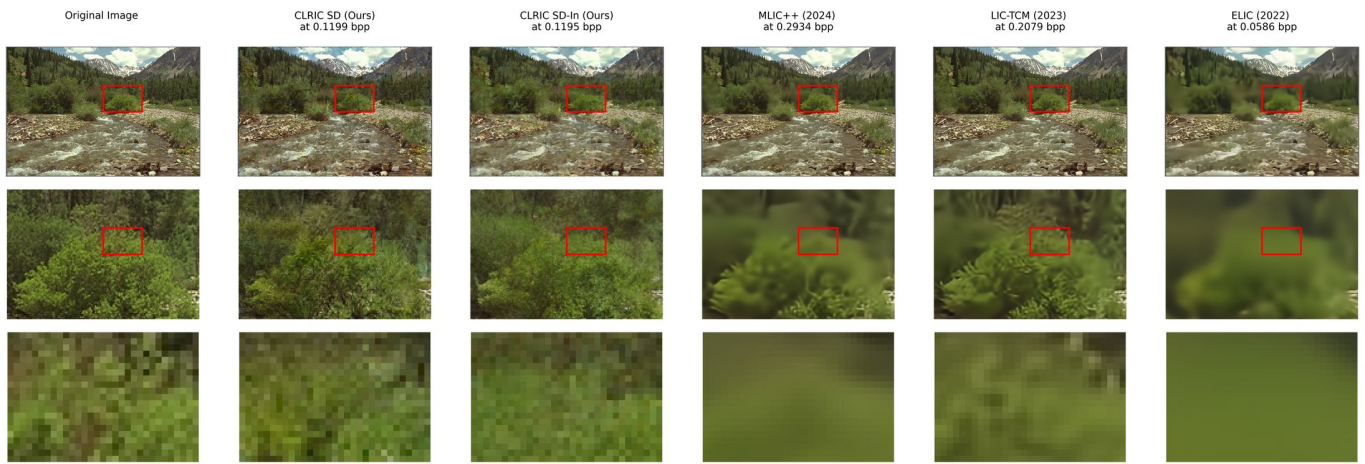}
\caption{Comparison of our approach against various models on image number 13 from the Kodak dataset.}
\label{fig:0.05_Kodak_13_bitrate_zoom_400}
\end{figure*}

\begin{figure*}[ht]
\centering
\includegraphics[width=\textwidth]{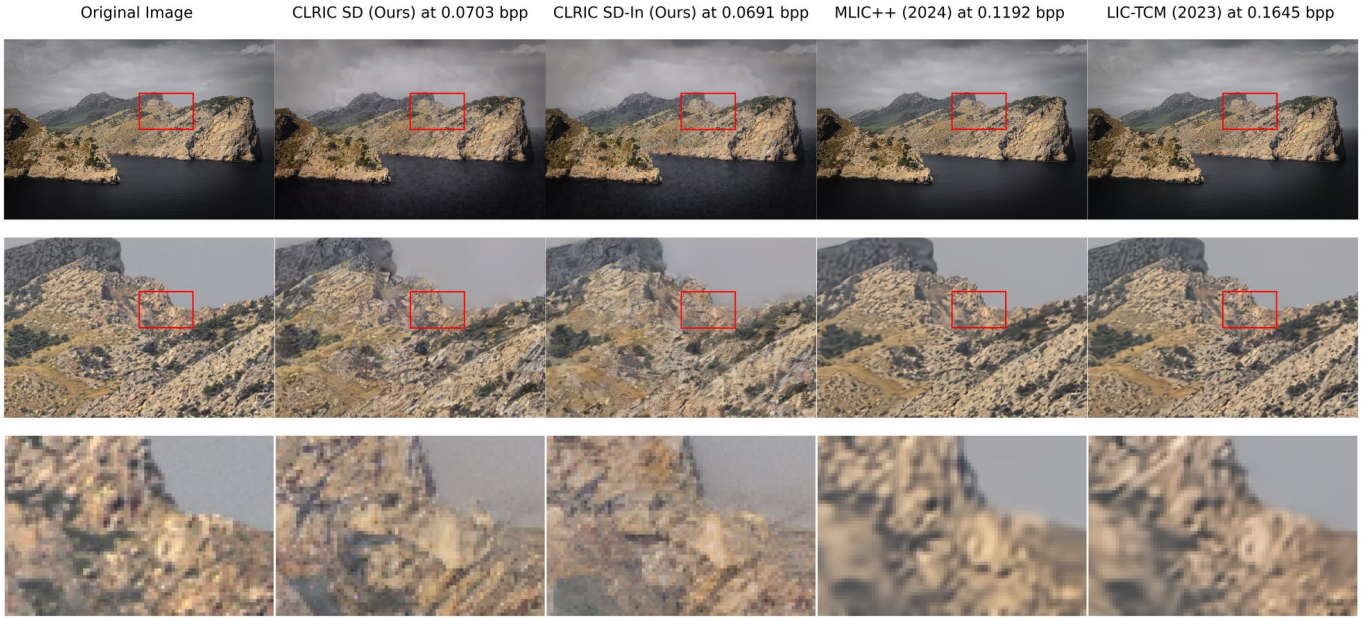}
\caption{Comparison of our approach against various models on image number 39 from the CLIC Professional Valid 2020 dataset.}
\label{fig:0.05_CLIC2020_39_bitrate_zoom_300}
\end{figure*}

\begin{figure*}[ht]
\centering
\includegraphics[width=\textwidth]{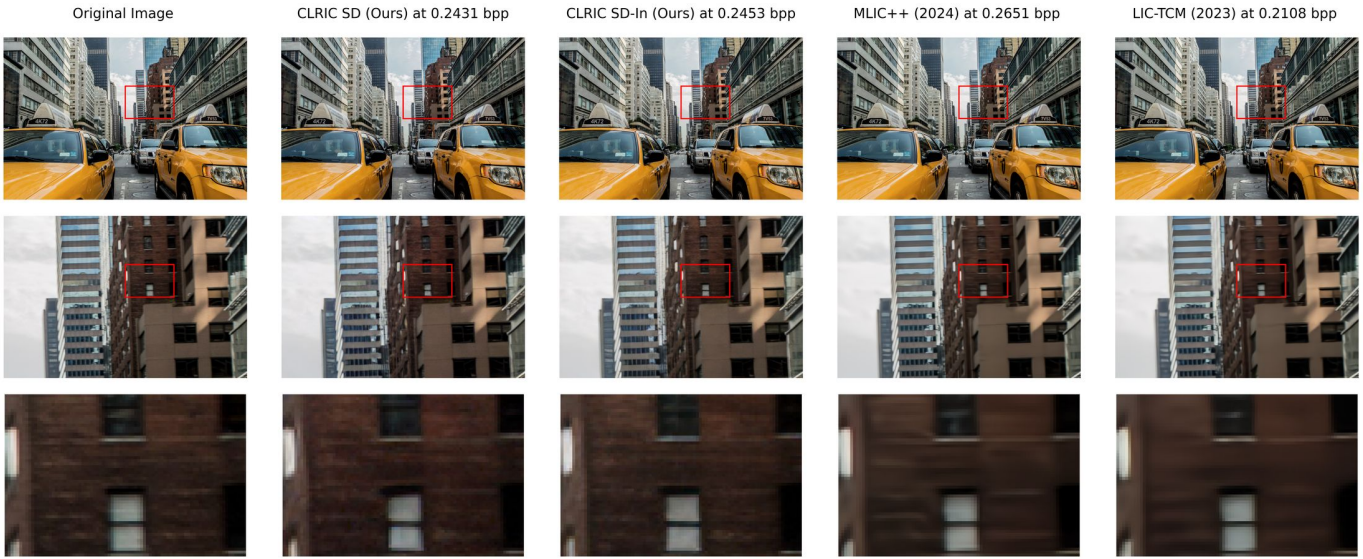}
\caption{Comparison of our approach against various models on image number 6 from the CLIC Professional Valid 2020 dataset.}
\label{fig:0.2_CLIC2020_6_bitrate_zoom_300}
\end{figure*}

\begin{figure*}[ht]
\centering
\includegraphics[width=\textwidth]{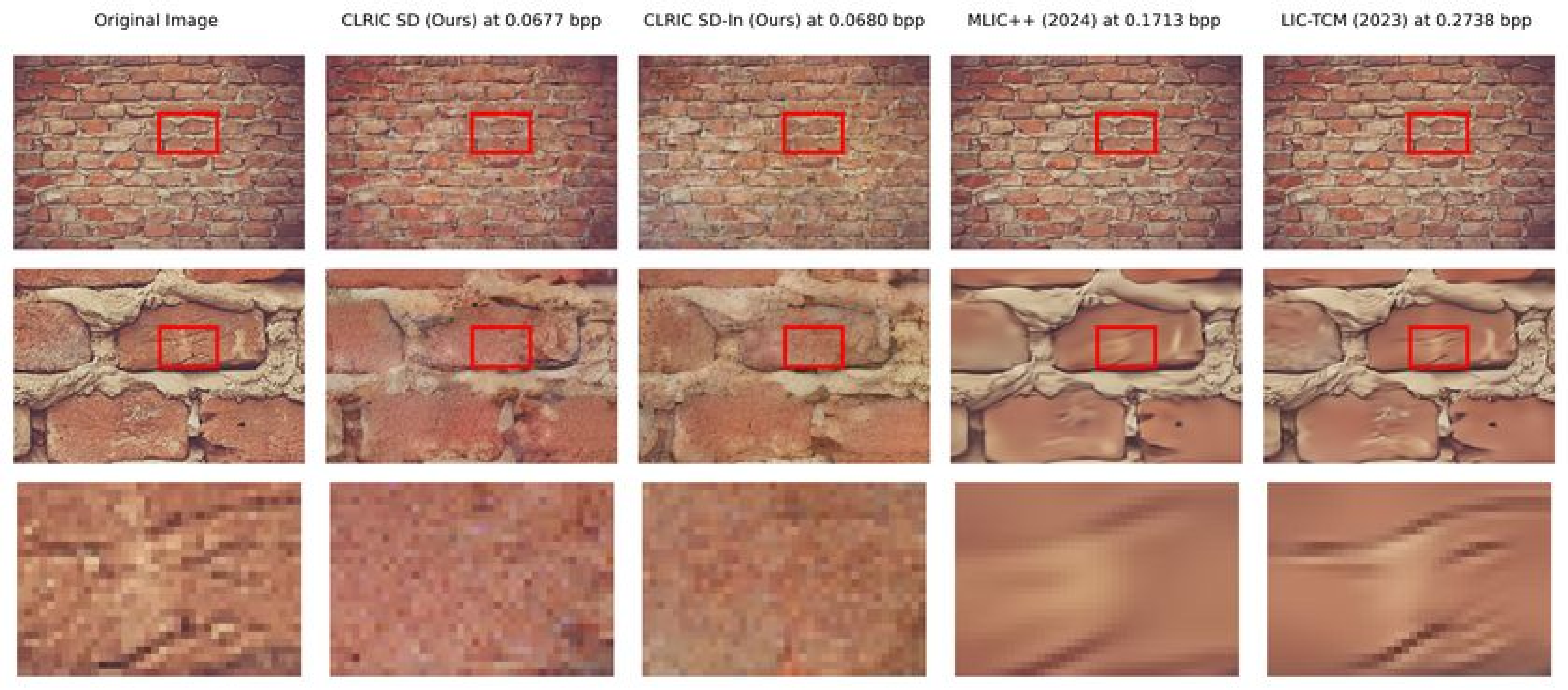}
\caption{Comparison of our approach against various models on image number 21 from the CLIC Professional Valid 2020 dataset.}
\label{fig:0.05_CLIC2020_21_bitrate_zoom_300}
\end{figure*}

\begin{figure*}[ht]
\centering
\includegraphics[width=\textwidth]{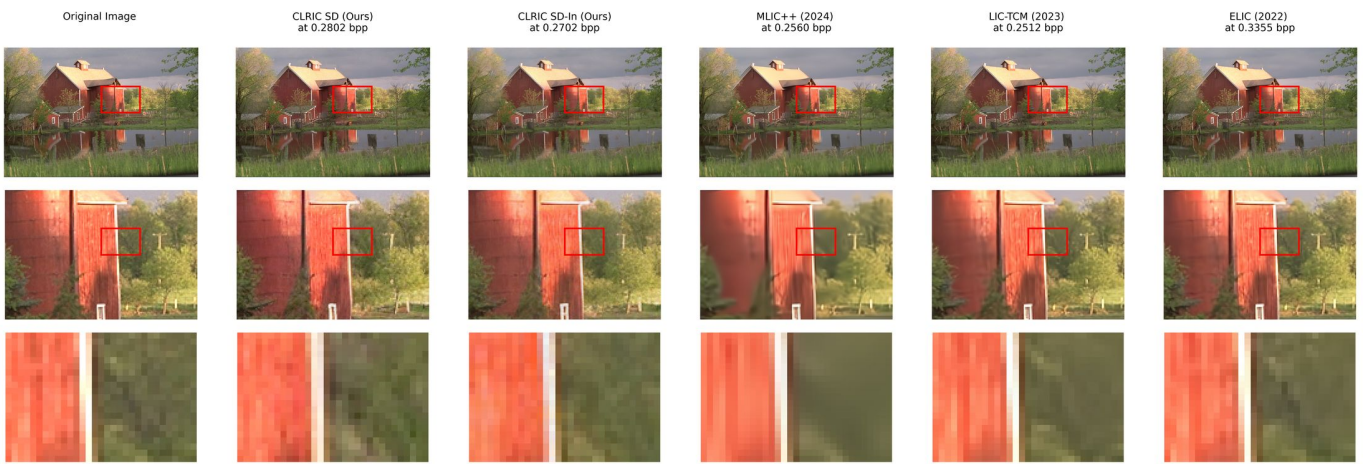}
\caption{Comparison of our approach against various models on image number 22 from the Kodak dataset.}
\label{fig:0.2_Kodak_22_bitrate_zoom_400}
\end{figure*}

\end{document}